\documentclass[aps,prb,preprint,showpacs,showkeys,floatfix]{revtex4}

\usepackage{graphicx,color}
\usepackage{multirow,slashbox}

\graphicspath{{figs/}}
\begin{document}
	
\title{Artificial Intelligence for Instability in Inorganic Perovskites: From Mechanism Discovery to Engineering Strategies}
	
\author{Xue Zhao}
\affiliation{Shanghai Key Laboratory of Mechanics in Energy Engineering, Shanghai Institute of Applied Mathematics and Mechanics, Shanghai Frontier Science Center of Mechanoinformatics, School of Mechanics and Engineering Science, Shanghai University, Shanghai 200072, People’s Republic of China}
	
\author{Chuan-Xin Cui}
\affiliation{Shanghai Key Laboratory of Mechanics in Energy Engineering, Shanghai Institute of Applied Mathematics and Mechanics, Shanghai Frontier Science Center of Mechanoinformatics, School of Mechanics and Engineering Science, Shanghai University, Shanghai 200072, People’s Republic of China}
	
\author{Zi-Hao Xu}
\affiliation{Shanghai Key Laboratory of Mechanics in Energy Engineering, Shanghai Institute of Applied Mathematics and Mechanics, Shanghai Frontier Science Center of Mechanoinformatics, School of Mechanics and Engineering Science, Shanghai University, Shanghai 200072, People’s Republic of China}
	
\author{Yuan-Long Pang}
\affiliation{Shanghai Key Laboratory of Mechanics in Energy Engineering, Shanghai Institute of Applied Mathematics and Mechanics, Shanghai Frontier Science Center of Mechanoinformatics, School of Mechanics and Engineering Science, Shanghai University, Shanghai 200072, People’s Republic of China}
	
\author{Jun-Jie Li}
\affiliation{Shanghai Key Laboratory of Mechanics in Energy Engineering, Shanghai Institute of Applied Mathematics and Mechanics, Shanghai Frontier Science Center of Mechanoinformatics, School of Mechanics and Engineering Science, Shanghai University, Shanghai 200072, People’s Republic of China}
	
\author{Jin-Wu Jiang}
\altaffiliation{Corresponding author: jiangjinwu@shu.edu.cn, jwjiang5918@hotmail.com}
\affiliation{Shanghai Key Laboratory of Mechanics in Energy Engineering, Shanghai Institute of Applied Mathematics and Mechanics, Shanghai Frontier Science Center of Mechanoinformatics, School of Mechanics and Engineering Science, Shanghai University, Shanghai 200072, People’s Republic of China}
	
\date{\today}
	
\begin{abstract}
Three-dimensional all-inorganic halide perovskites, represented by CsPbX$_3$ (X = Cl, Br, I), have attracted broad interest in photovoltaics, photodetectors, and light-emitting devices because of their outstanding optoelectronic properties. Their practical deployment, however, remains limited by instability under thermal, chemical, optical, and electrical stress. Conventional studies have established important experimental and theoretical foundations, but they still struggle with multimodal data, coupled degradation pathways, protocol dependence, sparse statistics, and uncertainty quantification. Artificial intelligence (AI) offers a practical route to address these limitations. This review summarizes recent progress in AI-assisted studies of instability in 3D CsPbX$_3$ and organizes the discussion around four linked tasks, including stability discrimination and diagnosis, microscopic mechanism analysis, consequence and reliability modeling, and engineering stability enhancement. For diagnosis, AI improves multimodal fusion, weak-signal extraction, early warning, and failure-mode attribution. For mechanistic analysis, it enables interpretable feature extraction, accelerated atomistic simulation, rare-event sampling, and cross-scale parameterization. For reliability studies, it supports degradation forecasting, lifetime-distribution modeling, risk estimation, and uncertainty-aware extrapolation. For engineering design, it helps transform empirical trial-and-error into closed-loop, multi-objective optimization under constraints of efficiency, manufacturability, and reproducibility. We further discuss the main limitations of current methods, especially in data quality, protocol consistency, benchmark design, interpretability, and transferability across domains. Finally, we outline future directions for the field, including standardized data infrastructures, interpretable cross-scale models, and tighter integration of AI with automated experiments and physics-based modeling. The aim of this review is to provide a coherent and practically useful framework for researchers seeking to use AI to understand, predict, and mitigate instability in inorganic perovskites.
\end{abstract}
	
	
\maketitle
\pagebreak
	
\section{Introduction}
	
Three-dimensional all-inorganic halide perovskites, especially CsPbX$_3$ (X = Cl, Br, I), have emerged as important materials for photovoltaics, photodetectors, and light-emitting devices because of their strong light absorption, tunable bandgaps, high defect tolerance, and solution-process compatibility.\cite{Protesescu2015CsPbX3,Zhang2018AllInorganicNCs,Jena2019ChemRev,Han2022Roadmap,Xiang2021InorganicStabilityReview,Wang2018SolventControlledGrowth,Wang2018CsPbI3Beyond15} Despite rapid gains in device performance, instability remains the main obstacle to practical deployment.\cite{Swarnkar2016AlphaCsPbI3,Ouedraogo2020NanoEnergy,Zhou2021DefectActivity,Wang2024CsPbX3Stability,Boyd2019DegradationMechanisms} Under heat, moisture, oxygen, illumination, and electrical bias, these materials may undergo phase transition, ion migration, defect evolution, interfacial reaction, and irreversible structural degradation.\cite{Ouedraogo2020NanoEnergy,Zhang2022DegradationPathways,Zhou2021DefectActivity,Wang2024CsPbX3Stability,Yuan2016IonMigrationAccChemRes,Eames2015IonicTransport} These processes interact across multiple length and time scales, so instability is not simply a matter of monotonic performance decay.\cite{Zhang2022DegradationPathways,Srivastava2023Spectroscopy,Wang2022VoidsStability}

Conventional stability research has produced valuable experimental and theoretical insight, but several structural difficulties remain. First, degradation pathways are strongly protocol-dependent, so conclusions drawn under one stress condition may not transfer directly to another.\cite{Zhang2022DegradationPathways,Wang2024CsPbX3Stability} Second, different observables, including diffraction, spectroscopy, microscopy, and device characteristics, are often analyzed separately even though they reflect the same evolving degradation process.\cite{Srivastava2023Spectroscopy,Zhang2022DegradationPathways} Third, stability datasets are commonly sparse, heterogeneous, and statistically limited, which complicates comparison, extrapolation, and uncertainty quantification.\cite{Zhang2022DegradationPathways,delaAsuncionNadal2025} These limitations have motivated growing interest in artificial intelligence (AI) as a tool for organizing data, improving diagnosis, extracting mechanisms, modeling reliability, and guiding optimization.\cite{Hartono2020CappingLayer,Kim2024AllInorganicML,Wang2024CompatibleMolecules,delaAsuncionNadal2025,Laufer2025ProcessMonitoring,Tao2021MLPerovskiteDesign,CorreaBaena2018AutomationMLHPC}

The value of AI in this context is not that it replaces experiment or physics-based modeling. Its value lies in its ability to integrate heterogeneous observations, detect weak and early signals, accelerate mechanistic analysis, and support reliability-aware engineering decisions under uncertainty.\cite{Hartono2020CappingLayer,Kim2024AllInorganicML,Wang2024CompatibleMolecules,Laufer2025ProcessMonitoring,delaAsuncionNadal2025} When coupled to physical constraints and careful validation, AI can help convert fragmented observations into a more coherent and predictive framework for instability research.\cite{Hartono2020CappingLayer,delaAsuncionNadal2025,Laufer2025ProcessMonitoring}

This review focuses on AI-assisted studies of instability in 3D CsPbX$_3$. Section 2 defines stability from thermodynamic, kinetic, and operational perspectives and clarifies the role of AI across the full workflow. Section 3 discusses stability discrimination and diagnosis, with emphasis on multimodal evidence, early warning, and probabilistic attribution. Section 4 addresses microscopic mechanisms of instability, including interpretable feature extraction, accelerated simulations, and cross-scale mechanistic modeling. Section 5 examines the consequences of instability for degradation, lifetime, and reliability. Section 6 then considers stability enhancement as a constrained engineering problem and discusses how AI can support closed-loop optimization. Section 7 focuses on data, benchmarks, reproducibility, and the practical limits of AI methods. The final section summarizes the main conclusions and outlines future directions.

Our aim is not simply to list AI methods that have been applied to perovskites. Instead, we seek to show how AI can be embedded into a coherent workflow for diagnosing instability, uncovering mechanisms, modeling reliability, and guiding engineering strategies. This review is thus intended less as a catalog of algorithms than as a framework for physically meaningful and practically useful AI in inorganic-perovskite stability research.
	
\section{Basic Knowledge of Stability}

\subsection{Three-Level Stability}

In the literature on CsPbX$_3$, the term stability is often used in a broad and sometimes ambiguous way.\cite{Ouedraogo2020NanoEnergy,Wang2024CsPbX3Stability,Xiang2021InorganicStabilityReview} In some cases, it refers to whether the material remains in the target crystal phase. In others, it describes how slowly the material evolves under external stress. In device studies, it usually denotes the retention of functional performance over time. These meanings are related, but they are not interchangeable. A material can be thermodynamically metastable yet appear stable within a finite time window if the kinetic barriers are sufficiently high.\cite{Swarnkar2016AlphaCsPbI3,Ouedraogo2020NanoEnergy,Sutton2018BlackPhaseStructure} Conversely, a material phase may remain intact while the device still degrades rapidly because of ion migration, interfacial reactions, or charge accumulation.\cite{Zhou2021DefectActivity,Zhang2022DegradationPathways,Yuan2016IonMigrationAccChemRes,Eames2015IonicTransport} This distinction is especially important for CsPbX$_3$, where soft lattices, strong ionicity, mobile defects, and polymorphic phase transitions are often strongly coupled.\cite{Ouedraogo2020NanoEnergy,Zhou2021DefectActivity,Wang2024CsPbX3Stability,Lee2022ACationScience,Xiang2021InorganicStabilityReview}

\begin{figure}[tb]
	\begin{center}
		\scalebox{1.8}[1.8]{\includegraphics[width=8cm]{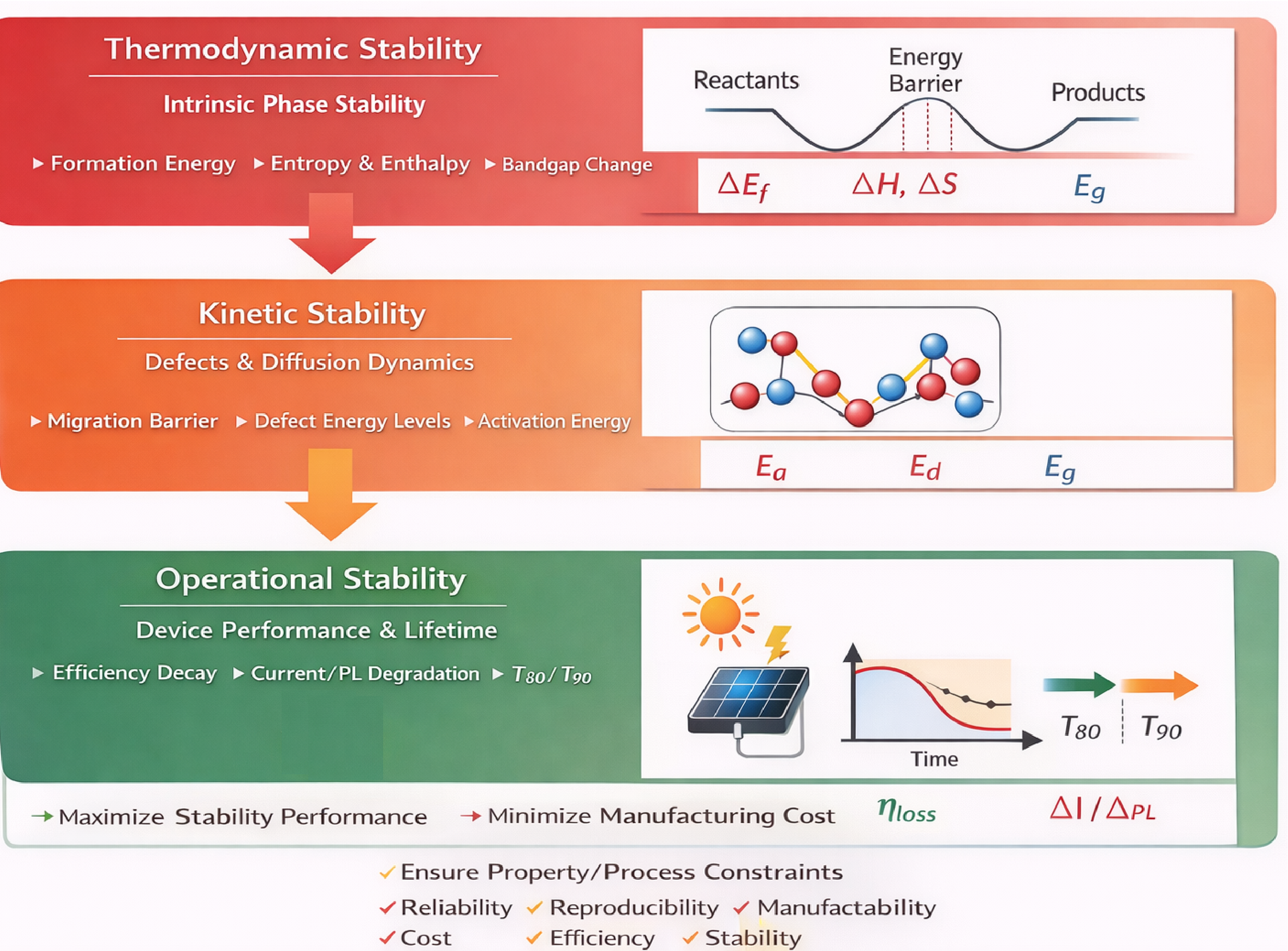}}
	\end{center}
	\caption{(Color online) Conceptual diagram of the three-level stability framework for CsPbX$_3$, including thermodynamic, kinetic, and operational stability, together with their typical observables and coupling relationships.}
	\label{fig_stability}
\end{figure}

\begin{figure}[tb]
	\begin{center}
		\scalebox{1.2}[1.2]{\includegraphics[width=8cm]{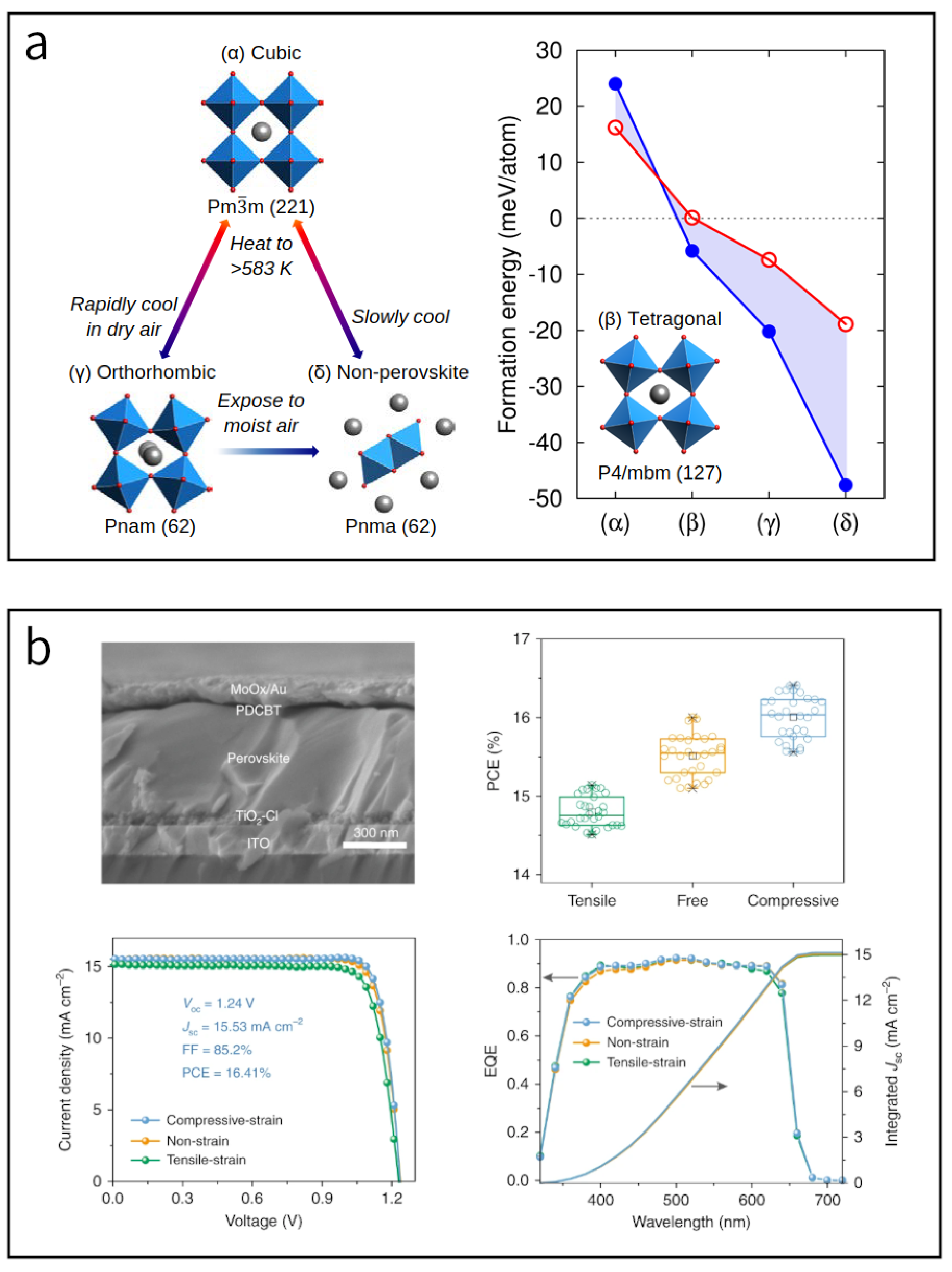}}
	\end{center}
	\caption{(Color online) Representative instability types at different levels. (a) Thermodynamic stability for CsPbX$_3$ at different temperature is governed by the energy barrier between different phases (Copyright © 2018 American Chemical Society).\cite{Sutton2018BlackPhaseStructure} (b) Kinetic stability of CsPbX$_3$ is enhanced by compressive strain, with improved performance.\cite{Xue2020StrainTransportLayers}}
	\label{fig_stability_real}
\end{figure}

This ambiguity makes direct comparison across studies difficult. Even for nominally similar compositions, different stress conditions, such as temperature, humidity, oxygen content, light intensity, or electrical bias, can activate different degradation pathways.\cite{Ouedraogo2020NanoEnergy,Zhang2022DegradationPathways} The chosen endpoint also matters. One study may define instability by the appearance of a diffraction peak, another by photoluminescence loss, and another by efficiency decay.\cite{Zhang2022DegradationPathways,delaAsuncionNadal2025} Apparently conflicting conclusions therefore often arise from different definitions rather than from genuine scientific disagreement. A unified framework is needed before mechanisms, lifetime, or stabilization strategies can be discussed in a consistent way.

In this review, stability is divided into three levels. Thermodynamic stability describes whether the target phase is energetically favored under given chemical and thermal conditions, or whether decomposition is thermodynamically driven.\cite{Swarnkar2016AlphaCsPbI3,Ouedraogo2020NanoEnergy,Wang2024CsPbX3Stability,Sutton2018BlackPhaseStructure,cui2023nonlinear} It answers the question of which state the material tends to adopt at equilibrium, but it does not determine whether that transformation will occur on a practical timescale. Kinetic stability concerns the rate of phase transition, decomposition, ion migration, defect evolution, and interfacial reaction under external stress.\cite{Zhou2021DefectActivity,Zhang2022DegradationPathways,Yuan2016IonMigrationAccChemRes,Eames2015IonicTransport,Xue2020StrainTransportLayers} It is controlled by barriers and rate processes, and it answers how fast degradation proceeds. Operational stability refers to the ability of a material or device to maintain its function under realistic working conditions.\cite{Ouedraogo2020NanoEnergy,jeong2020stable,Wang2024CsPbX3Stability,Boyd2019DegradationMechanisms} This is a system-level property that depends not only on the intrinsic material but also on microstructure, interfaces, transport layers, encapsulation, and test protocol. In this three-level framework, thermodynamics sets the driving force, kinetics controls the rate, and operational stability captures the final system-level consequence after these effects are coupled.

A schematic illustration of this framework is shown in Fig.~\ref{fig_stability}, while representative examples are shown in Fig.~\ref{fig_stability_real}. Thermodynamic stability sets the direction of evolution, kinetic stability governs how rapidly the system moves, and operational stability measures the practical outcome at the device level. This distinction is useful throughout the remainder of the review because it prevents mechanism, diagnosis, reliability, and engineering design from being collapsed into a single vague notion of stability.

\subsection{The Role of AI}

AI supports the full workflow of instability research, especially when data are heterogeneous, mechanisms are coupled, and experimental or computational resources are limited. Under these conditions, it can convert loosely connected observations into structured, comparable, and iteratively improvable knowledge.\cite{Hering2025AIAccelerated,delaAsuncionNadal2025,Laufer2025ProcessMonitoring}

At the data level, AI helps organize and standardize heterogeneous information.\cite{Maqsood2025InteroperablePerovskite,delaAsuncionNadal2025,Laufer2025ProcessMonitoring,Tao2021MLPerovskiteDesign,CorreaBaena2018AutomationMLHPC} Stability studies often contain large amounts of unstructured metadata, including sample history, stress conditions, encapsulation details, and measurement protocols. These details determine whether datasets are genuinely comparable.\cite{Maqsood2025InteroperablePerovskite,delaAsuncionNadal2025} AI can assist in protocol parsing, metadata extraction, unit normalization, and anomaly screening, thereby reducing the risk of comparing results that appear similar but were obtained under fundamentally different conditions.

At the diagnosis level, AI supports failure-mode recognition, weak-signal detection, and early warning.\cite{Hartono2020CappingLayer,Hering2025AIAccelerated,Laufer2025ProcessMonitoring,LeCorre2021RecombinationML} In CsPbX$_3$, many degradation processes first appear as subtle changes in spectra, images, or electrical signals before any obvious macroscopic failure is observed.\cite{Zhang2022DegradationPathways,Zhou2021DefectActivity,Wang2022VoidsStability} AI can identify these hidden patterns, detect abnormal drift, and improve the transferability of diagnostic criteria across batches, laboratories, and test conditions.\cite{Hartono2020CappingLayer,Hering2025AIAccelerated,Laufer2025ProcessMonitoring,delaAsuncionNadal2025}

At the mechanism level, AI can extract physically meaningful descriptors from high-dimensional observations.\cite{Kim2024AllInorganicML,Wang2024CompatibleMolecules,delaAsuncionNadal2025,Bian2026MLPotentials} From structural data, dynamic trajectories, or latent representations, it can reveal key local environments, probable reaction coordinates, or critical interfacial motifs. These outputs can then be linked to testable hypotheses connecting structure, barriers, rate processes, and macroscopic degradation behavior.\cite{Kim2024AllInorganicML,Wang2024CompatibleMolecules,Lee2022ACationScience}

At the reliability level, AI can model degradation trajectories, estimate lifetime distributions, and quantify risk under uncertainty.\cite{Hartono2020CappingLayer,Hering2025AIAccelerated,delaAsuncionNadal2025} For stability research, point predictions are rarely enough. Practical decisions require confidence intervals, failure probabilities, and an explicit measure of extrapolation risk. AI provides a framework for combining time-series behavior with statistical uncertainty in a way that is relevant for engineering decisions.\cite{Hering2025AIAccelerated,delaAsuncionNadal2025,Laufer2025ProcessMonitoring}

At the engineering level, AI enables multi-objective optimization of stability-improvement strategies. Stabilizing CsPbX$_3$ usually requires balancing performance, durability, cost, manufacturability, and reproducibility. AI can search this constrained design space more efficiently, especially when combined with active learning or closed-loop experiments.\cite{Kim2024AllInorganicML,Wang2024CompatibleMolecules,Hering2025AIAccelerated,Laufer2025ProcessMonitoring,CorreaBaena2018AutomationMLHPC}
	
\section{Stability Discrimination and Diagnosis}

Having defined the three-level stability framework in Section 2, the next step is to determine what kind of instability is occurring in a given CsPbX$_3$ system and how early it can be recognized. This section therefore focuses on stability discrimination and diagnosis under explicitly defined stress protocols and evaluation endpoints. The central question is which process is dominant, where it initiates, how it propagates, and whether its onset can be detected before visible failure emerges. In practice, diagnosis should identify the dominant degradation mode, its spatial distribution, its temporal evolution, and the relative likelihood of competing mechanisms rather than forcing all observations into a single-cause narrative.

\subsection{Problem Definition}

\subsubsection{Instability Processes}

In 3D CsPbX$_3$, stability discrimination usually involves several coupled degradation processes. These include phase instability, such as polymorphic transitions between black and yellow phases, stress-induced phase evolution, and grain rearrangement.\cite{Swarnkar2016AlphaCsPbI3,Ouedraogo2020NanoEnergy,Sutton2018BlackPhaseStructure,Wang2018CsPbI3Beyond15} Chemical degradation is another major pathway, especially reactions with moisture and oxygen that often begin at surfaces and grain boundaries.\cite{Ouedraogo2020NanoEnergy,Zhang2022DegradationPathways,zhao2026charge} Ion migration and compositional redistribution are also common, particularly under coupled light, heat, and electric fields, where halide vacancies, halide ions, and mixed-halide segregation generate local compositional and spectral heterogeneity.\cite{Zhou2021DefectActivity,Chen2016OriginJVHysteresis,Yuan2016IonMigrationAccChemRes,Eames2015IonicTransport} Defect states may evolve through generation, aggregation, or passivation failure, while interfaces can degrade through reactions with transport layers and electrodes, ion accumulation, and changes in contact properties.\cite{Zhou2021DefectActivity,Zhang2022DegradationPathways} Microstructural evolution, including grain coarsening, crack formation, void growth, and phase-boundary motion, further complicates diagnosis.\cite{Zhang2022DegradationPathways,Srivastava2023Spectroscopy,Wang2022VoidsStability}

\subsubsection{Observational Signals}

The diagnosis of instability in CsPbX$_3$ relies on multimodal observations. Structural information is commonly obtained from X-ray diffraction, grazing-incidence wide-angle X-ray scattering, and Raman spectroscopy, while microstructural evolution is monitored by scanning electron microscopy, transmission electron microscopy, atomic force microscopy, and optical microscopy.\cite{Srivastava2023Spectroscopy,Zhang2022DegradationPathways} Chemical composition and depth profiles are often probed by X-ray photoelectron spectroscopy and time-of-flight secondary ion mass spectrometry.\cite{Srivastava2023Spectroscopy} Optical measurements, including absorption, steady-state photoluminescence, time-resolved photoluminescence, and spectral shifts or broadening, capture electronic and excitonic responses.\cite{Srivastava2023Spectroscopy} At the device level, time-dependent current density--voltage curves, hysteresis, impedance spectra, transient electrical responses, and noise signals provide additional evidence.\cite{Sanchez2014SlowDynamicProcesses,Chen2016OriginJVHysteresis,Wang2024HysteresisDiagnosis} Under stress, these measurements form correlated time series rather than isolated descriptors.

\subsubsection{Evaluation Metrics}

Traditional stability evaluation is often based on endpoint criteria, such as the appearance of a new diffraction peak, photoluminescence decay to a threshold, or efficiency loss to a prescribed fraction of the initial value. For diagnosis, however, process-level metrics are more informative.\cite{Hering2025AIAccelerated,Wang2024HysteresisDiagnosis} Useful indicators include the magnitude and rate of signal drift, the time at which inflection points emerge, anomaly scores relative to baseline behavior, and spatial statistics describing the extent and connectivity of degraded regions.\cite{Ji2023SelfSupervisedDegradation,Wang2024HysteresisDiagnosis} Attribution metrics are also important because they estimate the relative contribution of competing failure modes instead of forcing a single deterministic label.\cite{Wang2024HysteresisDiagnosis} Because diagnostic conclusions are often used to guide subsequent experiments or engineering decisions, uncertainty estimates should accompany all major outputs.\cite{delaAsuncionNadal2025,Hering2025AIAccelerated}

\subsection{Conventional Approaches for Stability Discrimination and Diagnosis}

Before the widespread use of AI, stability discrimination in CsPbX$_3$ mainly relied on a small number of characterization tools interpreted through human experience.\cite{Sanchez2014SlowDynamicProcesses,Chen2016OriginJVHysteresis,Zhang2022DegradationPathways,Boyd2019DegradationMechanisms} A common workflow was to monitor X-ray diffraction, photoluminescence, or device performance during aging tests, identify characteristic changes using empirical thresholds, and then perform post-hoc attribution with limited compositional or interfacial analysis.\cite{Zhang2022DegradationPathways,Wang2024CsPbX3Stability} This approach is intuitive and easy to align with conventional materials science narratives, but it is strongly dependent on expert judgment. It is also poorly suited to separating coupled degradation pathways, detecting weak early-stage signals, or transferring across laboratories and testing protocols.\cite{Habisreutinger2018HysteresisIndex,Zhang2022DegradationPathways} Yao \textit{et al.} used in situ optical microscopy, cathodoluminescence, and X-ray diffraction to follow humidity-induced degradation in all-inorganic CsPbI$_2$Br films.\cite{Yao2022HumidityDegradation} They showed that the degradation route and its spatial heterogeneity depend strongly on relative humidity, with the photoactive $\alpha$ phase progressively converting into the nonperovskite $\delta$ phase through microscopically nonuniform pathways.\cite{Yao2022HumidityDegradation} This type of study illustrates the strength of conventional diagnosis. It directly ties visible microstructural and optical signatures to a specific stressor, but it still relies on manually chosen observables and expert interpretation across measurement channels.

More systematic non-learning methods, such as principal component analysis, clustering, and simple regression, improved repeatability to some extent by reducing dimensionality and extracting low-dimensional descriptors.\cite{Odabasi2020StabilityML,Shaji2025HysteresisStability} Even so, these approaches still depended heavily on manual feature design, used limited multimodal integration, and struggled with domain shifts caused by differences in preparation, testing, and instrumentation. As a result, traditional diagnosis was largely confirmatory. It could often determine what happened after failure, but it was much less effective at identifying when degradation began, why it started, or how it would evolve.\cite{Hering2025AIAccelerated,delaAsuncionNadal2025}

\subsection{Core Challenges in Stability Discrimination and Diagnosis}

The difficulty of stability diagnosis lies in building a transferable framework for early warning, attribution, and prediction.\cite{Hering2025AIAccelerated,delaAsuncionNadal2025} Early warning is challenging because degradation often starts long before a clear macroscopic endpoint appears. Subtle photoluminescence broadening, local lifetime shortening, small changes in impedance, or weak texture variations in images can easily be buried under noise and sample heterogeneity.\cite{Srivastava2023Spectroscopy,Ji2023SelfSupervisedDegradation} Attribution is equally difficult because phase transitions, ion migration, interfacial reactions, defect evolution, and environmental attack are often strongly coupled.\cite{Zhou2021DefectActivity,Zhang2022DegradationPathways,Wang2024HysteresisDiagnosis} The same macroscopic signal may arise from different dominant mechanisms, while a single mechanism may appear differently across measurement channels. Without multimodal integration, attribution easily degenerates into a plausible but weakly testable narrative.\cite{Hering2025AIAccelerated,Wang2024HysteresisDiagnosis} A third challenge is cross-protocol transferability. Variations in laboratory conditions, device structures, sample batches, and stress combinations introduce major domain shifts, so rules derived under one protocol may fail under another.\cite{Shaji2025HysteresisStability,delaAsuncionNadal2025}

\subsection{AI-enabled Approaches for Stability Discrimination and Diagnosis}

AI changes stability diagnosis by shifting the focus from post-hoc identification to process-aware, multimodal, and probabilistic inference.\cite{Ji2023SelfSupervisedDegradation,Wang2024HysteresisDiagnosis,Hering2025AIAccelerated,delaAsuncionNadal2025,Tao2021MLPerovskiteDesign} The key issue is how the diagnostic problem is formulated. For example, what data are used as input, what outputs are required, how stress protocols are encoded, and how uncertainty is propagated into the final decision.\cite{Hering2025AIAccelerated,delaAsuncionNadal2025}  Wang \textit{et al.} developed a device-diagnosis model, which used hysteresis features in combination with an ion-incorporated drift--diffusion framework to diagnose perovskite solar-cell deficiencies and degradation states.\cite{Wang2024HysteresisDiagnosis} It was demonstrated that a readily measured J-V response can be mapped back to interpretable device-physics causes such as interfacial extraction imbalance, trap-assisted loss, and ion-related degradation signatures.\cite{Wang2024HysteresisDiagnosis} Ji \textit{et al.} provided a complementary imaging-based example by using self-supervised deep learning to denoise and analyze multispectral operando data, thereby tracking the spatial evolution of degradation in perovskite light-emitting diodes before obvious macroscopic failure occurred.\cite{Ji2023SelfSupervisedDegradation}

\begin{figure}[tb]
	\begin{center}
		\scalebox{1.8}[1.8]{\includegraphics[width=8cm]{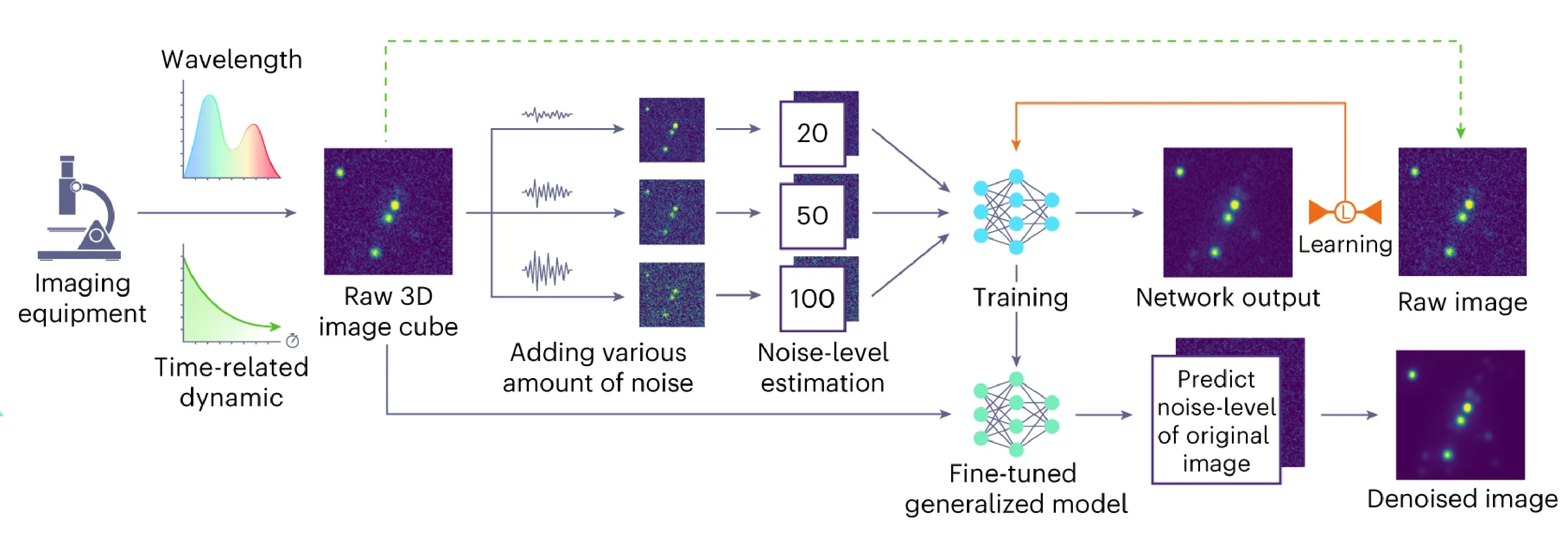}}
	\end{center}
	\caption{(Color online) A convolutional residual network with a universal noise-level estimator for the hyperspectral datasets.\cite{Ji2023SelfSupervisedDegradation}}
	\label{fig_ai-diagnosis}
\end{figure}

The main advantage of multimodal fusion is that different degradation mechanisms leave different signatures in different channels.\cite{Ji2023SelfSupervisedDegradation,Laufer2025ProcessMonitoring} Microscopy images, spectral data, and electrical time series can be projected into a shared representation, from which the model can infer degradation type, severity, or spatial distribution. Compared with single-signal threshold methods, this strategy is better suited to separating mechanisms that converge to a similar endpoint.\cite{Wang2024HysteresisDiagnosis} An example is the self-supervised deep-learning framework shown in Fig.~\ref{fig_ai-diagnosis}, which was used to track degradation in perovskite light-emitting diodes through multispectral imaging.\cite{Ji2023SelfSupervisedDegradation}

For scanning electron microscopy, optical microscopy, or atomic force microscopy data, AI can segment degraded phases, cracks, voids, and phase boundaries, detect the earliest abnormal regions, and quantify their spatial statistics as a function of stress time.\cite{Ji2023SelfSupervisedDegradation,Zhang2023MachineVisionGrains,Zhang2024GrainCharacteristics,Li2024GrainBoundaryDefects,Laufer2025ProcessMonitoring} Recent studies on perovskite micrographs have shown that machine-vision-based analysis can extract grain morphology and grain-boundary descriptors in a more reproducible and statistically meaningful way than manual inspection.\cite{Zhang2023MachineVisionGrains,Zhang2024GrainCharacteristics,Li2024GrainBoundaryDefects} This turns qualitative image inspection into reproducible spatial descriptors and helps establish whether degradation preferentially initiates at grain boundaries, surfaces, or interfaces.

For photoluminescence, absorption, Raman, and diffraction data, AI is particularly useful for weak-signal extraction and trajectory analysis.\cite{Ji2023SelfSupervisedDegradation,Srivastava2023OpticalBehavior,Marchenko2025XRDRecognition,Wang2024HysteresisDiagnosis} Recent work has shown that machine-learning models can capture subtle optical-response trajectories from in situ photoluminescence data and can also classify structurally similar perovskite phases directly from diffraction patterns.\cite{Srivastava2023OpticalBehavior,Marchenko2025XRDRecognition} Early degradation often appears as subtle shifts in peak position, width, intensity, or background long before obvious spectral failure occurs.\cite{Srivastava2023Spectroscopy,Srivastava2023OpticalBehavior} Data-driven feature learning can therefore reveal degradation pathways that would be difficult to isolate using manual peak fitting alone.\cite{Ji2023SelfSupervisedDegradation,Srivastava2023OpticalBehavior,Marchenko2025XRDRecognition}

Because degradation is inherently dynamic, time-series modeling is central to stability diagnosis.\cite{Ji2023SelfSupervisedDegradation,Wang2024HysteresisDiagnosis,DunlapShohl2024OperationalLifetimes,Paraskeva2024OutdoorMonitoring,Kouroudis2024WaveletAidedPrediction,Chen2025StabilitySHAP} AI can track anomaly scores, infer the probability of approaching an inflection point, and predict short-term trends under ongoing stress.\cite{Ji2023SelfSupervisedDegradation,Wang2024HysteresisDiagnosis,DunlapShohl2024OperationalLifetimes} It can also use early-stage degradation trajectories to predict long-term operational behavior under accelerated or outdoor conditions.\cite{DunlapShohl2024OperationalLifetimes,Kouroudis2024WaveletAidedPrediction,Chen2025StabilitySHAP,Paraskeva2024OutdoorMonitoring} This allows failure assessment to move from a late-stage threshold criterion toward an earlier warning window, where the system can already be identified as deviating from its normal trajectory.\cite{Ji2023SelfSupervisedDegradation,Hering2025AIAccelerated,DunlapShohl2024OperationalLifetimes,Kouroudis2024WaveletAidedPrediction}

Under coupled degradation, it is important to estimate the probability that phase change, ion migration, interfacial reaction, or defect evolution is dominant under the available evidence.\cite{Wang2024HysteresisDiagnosis,Hering2025AIAccelerated} This can be improved by multitask learning, by incorporating physically motivated priors on signal sensitivity, and by explicitly allowing uncertain predictions when the evidence is insufficient. Diagnosis then becomes not only a classification tool but also a guide for selecting the next most informative experiment.\cite{Hering2025AIAccelerated,delaAsuncionNadal2025}

For diagnostic tools to be reusable, domain shifts must be treated explicitly.\cite{Shaji2025HysteresisStability,delaAsuncionNadal2025} Stress variables such as temperature, humidity, illumination, and bias should be incorporated as conditional inputs, so that the model becomes protocol-aware rather than protocol-blind. Domain adaptation and calibration are also essential for aligning outputs across laboratories and batches.\cite{Shaji2025HysteresisStability,Hering2025AIAccelerated} Equally important, training and test splits should be defined by batch, laboratory, or protocol rather than by random shuffling, otherwise reported accuracy may greatly overestimate practical transferability.\cite{delaAsuncionNadal2025,Hering2025AIAccelerated}

\subsection{Validation and Credibility}

The validation standard for stability diagnosis should be stricter than that used in routine machine-learning tasks because the outputs are intended to support mechanistic interpretation and engineering decisions. Internal cross-validation alone is insufficient, as it often conceals protocol- or batch-induced domain shifts. More convincing evidence comes from external validation across batches, laboratories, or stress protocols, together with prospective tests on newly prepared samples or newly designed degradation experiments.\cite{Shaji2025HysteresisStability,Wang2024HysteresisDiagnosis} For multimodal models, ablation analysis is necessary to verify that predictions are not dominated by shortcut features unrelated to degradation physics.\cite{Ji2023SelfSupervisedDegradation,Laufer2025ProcessMonitoring} Probability outputs should be calibrated, and uncertainty should be reported explicitly rather than treated as a secondary detail. Finally, all comparisons must be made under matched or carefully mapped stress protocols; otherwise, even a numerically accurate model may be answering a different diagnostic question from the one intended.\cite{Hering2025AIAccelerated,delaAsuncionNadal2025}
	
\section{Microscopic Mechanisms of Instability}

Section 3 addressed what instability looks like and how it can be diagnosed. This section turns to why does a given CsPbX$_3$ system follow one degradation pathway rather than another under a particular stress condition? The focus is therefore on the microscopic origins of instability under temperature, humidity or oxygen, illumination, electrical bias, and their coupling.\cite{Marronnier2018AnharmonicityDisorder,Chen2022KineticPathwayGammaDelta,Zhang2022DegradationPathways,Lee2022ACationScience,cui2024molecular,Liang2025BSiteDopingIonMigration} The goal is to connect the thermodynamic and kinetic concepts introduced in Section 2 with the failure modes, spatial initiation sites, and early-stage signatures discussed in Section 3. Only with that connection can diagnostic observations be translated into testable mechanistic hypotheses and into physically meaningful variables, such as energy barriers, defect populations, diffusion coefficients, and reaction rates, that later inform degradation and reliability analysis in Section 5.\cite{Hering2025AIAccelerated,delaAsuncionNadal2025}

\begin{figure}[tb]
	\begin{center}
		\scalebox{1.8}[1.8]{\includegraphics[width=8cm]{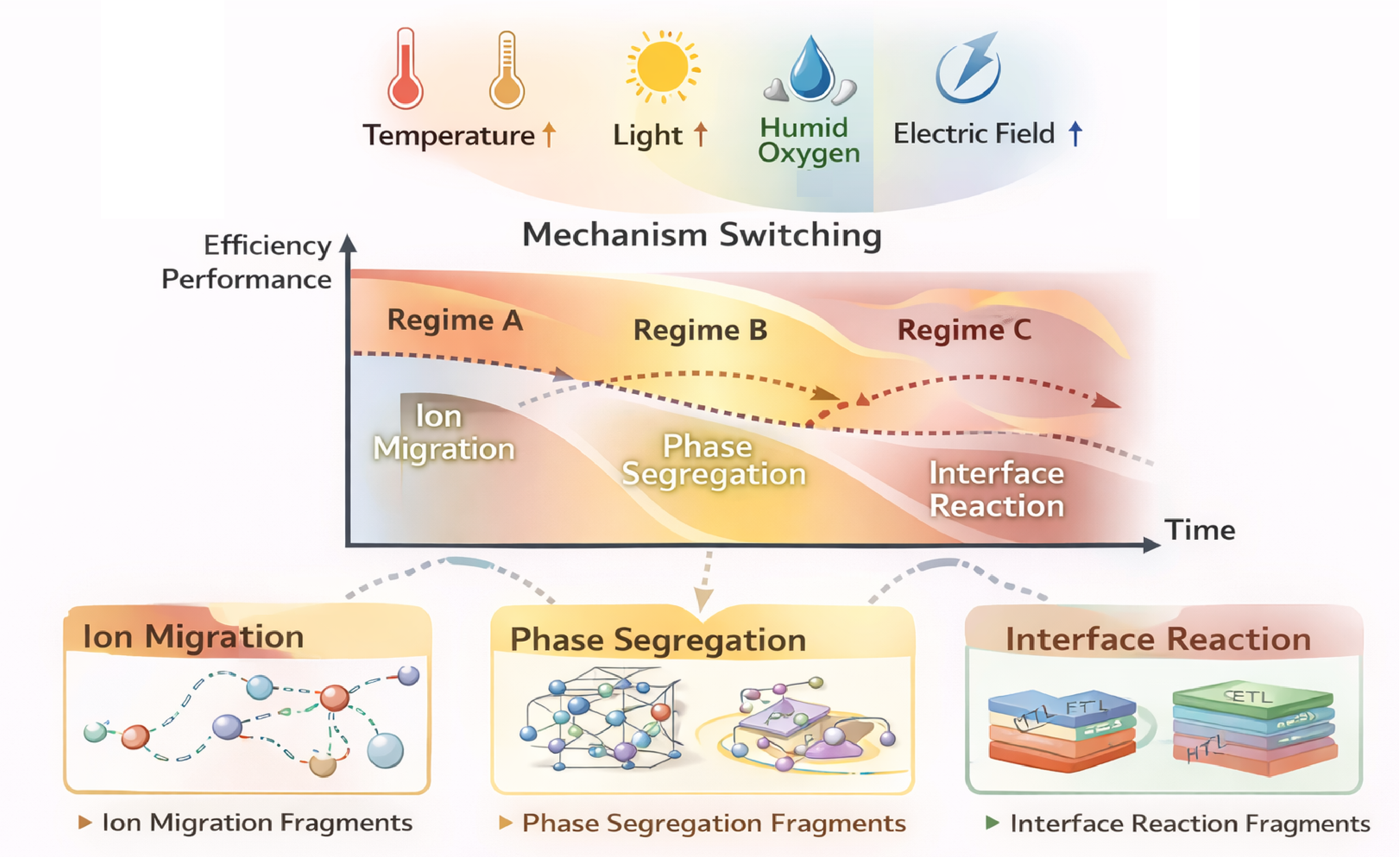}}
	\end{center}
	\caption{(Color online) Typical microscopic mechanisms for the instability of CsPbX$_3$ underlying different external conditions.}
	\label{fg_meichanism}
\end{figure}

\begin{figure}[tb]
	\begin{center}
		\scalebox{1.8}[1.8]{\includegraphics[width=8cm]{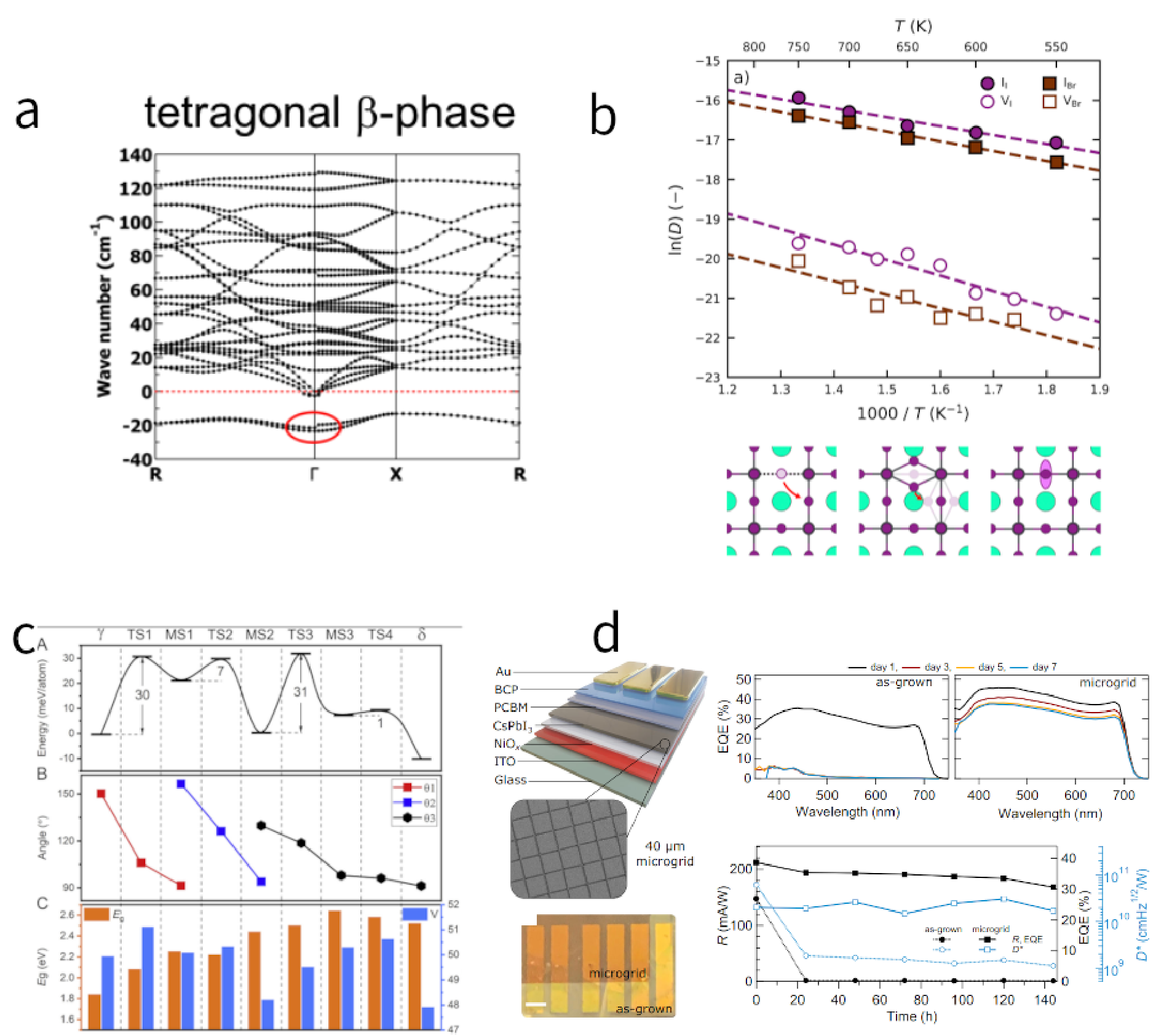}}
	\end{center}
	\caption{(Color online) Typical instability mechanisms. (a) Soft phonon modes are responsible for the phase instability according to the vibration morphology of the soft phonon mode (Copyright © 2018 American Chemical Society).\cite{Marronnier2018AnharmonicityDisorder} (b) Temperature dependent vacancy self-diffusion coefficient along different defect migration pathways.\cite{Pols2023DefectMigrationMLFF} (c) Energy barrier along major major phase transition pathways for CsPbX$_3$.\cite{Chen2022KineticPathwayGammaDelta} (d) The interface is important for the stability.\cite{Steele2022EmbeddedNetworkCsPbI3}}
	\label{fg_meichanism_real}
\end{figure}

\subsection{Problem Definition}

Microscopic mechanism analysis in CsPbX$_3$ involves two closely related aspects.\cite{Marronnier2018AnharmonicityDisorder,Chen2022KineticPathwayGammaDelta,Wang2024CsPbX3Stability} The first is thermodynamic driving force, namely the relative stability of the target phase with respect to competing phases, the tendency toward decomposition, and the preferred interfacial or surface configurations under given chemical and thermal conditions.\cite{Marronnier2018AnharmonicityDisorder,Ouedraogo2020NanoEnergy,Sutton2018BlackPhaseStructure,Steele2022EmbeddedNetworkCsPbI3} The second is kinetics, which determines the dominant pathways, intermediate states, and rates of phase transition, decomposition, ion migration, defect aggregation, and interfacial reaction.\cite{Chen2022KineticPathwayGammaDelta,Pols2023DefectMigrationMLFF,Tyagi2025TracingIonMigrationMLFF,Yuan2016IonMigrationAccChemRes,Eames2015IonicTransport,Liang2025BSiteDopingIonMigration} Thermodynamics determines what is favored, whereas kinetics determines when, where, and how fast the process occurs on a finite timescale.

A central difficulty is that mechanistic understanding must span multiple scales.\cite{Zhang2022DegradationPathways,Hering2025AIAccelerated} Atomic-scale events, such as local distortions, defect formation, or ion hopping, must be linked to nanoscale heterogeneity at grain boundaries, surfaces, and interfaces, and further connected to device-scale quantities such as electric-field distribution, temperature gradients, and carrier accumulation.\cite{Liu2023GrainBoundarySlidingMLFF,Samatov2024IonMigrationGBMLFF} Ultimately, these effects must explain observable changes in spectra, electrical response, and morphology. Without such a cross-scale connection, many mechanistic arguments remain plausible but not predictive.

For this reason, mechanism analysis in this review is judged by a stricter standard than simple post-hoc interpretation. A useful mechanistic statement should lead to falsifiable predictions.\cite{Hering2025AIAccelerated,delaAsuncionNadal2025} It should specify how observable quantities change when a control variable, such as temperature, bias, composition, interface material, or defect concentration, is varied. It should also identify the key microscopic quantities behind this response, including barriers, rate constants, and diffusion coefficients, as illustrated in Fig.~\ref{fg_meichanism}. Some representative instability mechanisms are shown in Fig.~\ref{fg_meichanism_real}. Just as importantly, it should clarify how the proposed mechanism can be distinguished from competing explanations in experiment or simulation.

\subsection{Conventional Approaches for Microscopic Mechanisms of Instability}

Before the widespread use of AI, mechanistic studies of CsPbX$_3$ mainly relied on first-principles calculations, molecular simulations, and multiscale modeling.\cite{Marronnier2018AnharmonicityDisorder,Chen2022KineticPathwayGammaDelta,Zhou2021DefectActivity} Density functional theory has been widely used to compare phase stability, calculate defect formation energies, determine migration barriers through methods such as the nudged elastic band approach, and analyze lattice instability through phonons or soft modes.\cite{Marronnier2018AnharmonicityDisorder,Chen2022KineticPathwayGammaDelta} Its strength lies in direct physical interpretability, but its application is limited by system size and by the difficulty of representing realistic grain boundaries, reconstructed surfaces, and complex interfaces. Lin \textit{et al.} combined controlled-humidity experiments with time-resolved photoluminescence and structural characterization to quantify the moisture-induced phase-transformation kinetics of CsPbI$_3$.\cite{Lin2021MoisturePhaseTransformation} They showed that moisture changes the transformation kinetics in a nontrivial way, reflecting a competition between adsorption-assisted destabilization and moisture desorption, rather than a simple monotonic acceleration of degradation.\cite{Lin2021MoisturePhaseTransformation}

Molecular dynamics, whether based on classical force fields or first-principles descriptions, has provided insight into lattice fluctuations, ion diffusion, and finite-temperature structural evolution.\cite{Marronnier2018AnharmonicityDisorder,Pols2023DefectMigrationMLFF,xu2026survey} Enhanced sampling methods further expand the accessible event space.\cite{Chen2022KineticPathwayGammaDelta} Even so, this approach remains constrained by the accuracy of the potential, particularly for reactions, charge transfer, polarization, and long-range electrostatics, as well as by the limited time scales accessible in practice.\cite{Pols2023DefectMigrationMLFF}

At a larger scale, kinetic Monte Carlo, rate equations, drift--diffusion models, and continuum descriptions have been used to interpret ion transport, hysteresis, and device-level degradation trends.\cite{Chen2016OriginJVHysteresis,Sanchez2014SlowDynamicProcesses} These methods are valuable because they bridge microscopic parameters and macroscopic response, but they depend strongly on the quality and transferability of the input parameters. They also tend to simplify spatial heterogeneity and the coupling between different degradation channels. As a result, traditional approaches have been effective for analyzing individual processes, but much less powerful for treating coupled mechanisms, heterogeneous structures, and long-time degradation under realistic operating conditions.\cite{Hering2025AIAccelerated}

\subsection{Core Challenges in Microscopic Mechanisms of Instability}

The main challenges of microscopic mechanism analysis in CsPbX$_3$ arise from the complexity of the local structure, the rarity of key events, and the strong coupling across scales and operating conditions.\cite{Zhou2021DefectActivity,Hering2025AIAccelerated} In many cases, the decisive local structural features are not known in advance. Processes such as phase nucleation, defect aggregation, ion hopping, and interfacial reaction are controlled by local coordination, strain, chemistry, and charge redistribution, especially near grain boundaries, surfaces, and interfaces.\cite{Liu2023GrainBoundarySlidingMLFF,Samatov2024IonMigrationGBMLFF,Xue2020StrainTransportLayers,pang2025strain,zhao2025surface} Conventional hand-crafted descriptors, such as bond lengths or bond angles, are often not sufficient to identify the dominant variables.

A second challenge is that many degradation processes are governed by rare events and long-time barriers.\cite{Chen2022KineticPathwayGammaDelta,Pols2023DefectMigrationMLFF,Tyagi2025TracingIonMigrationMLFF,Yuan2016IonMigrationAccChemRes,Eames2015IonicTransport,Liang2025BSiteDopingIonMigration} Ion migration, phase nucleation, and interfacial reactions may occur only occasionally, but once triggered they can drive irreversible evolution. Standard molecular dynamics is usually too short to capture reliable statistics for such processes, while first-principles methods typically provide only isolated barriers rather than the full event landscape in realistic defective structures.\cite{Pols2023DefectMigrationMLFF}

A third challenge is the lack of a robust cross-scale bridge. Section 3 may identify where degradation begins and how it evolves in time, but converting atomic-scale energies and barriers into macroscopic degradation behavior requires a richer framework involving event networks, defect densities, field distributions, and statistical variability.\cite{Hering2025AIAccelerated,delaAsuncionNadal2025} Without this bridge, mechanisms remain descriptive rather than predictive.

Finally, mechanistic conclusions are highly protocol-dependent. Illumination, bias, heat, and environment often act simultaneously, and their coupling can shift the dominant degradation pathway.\cite{Zhang2022DegradationPathways,Zhou2021DefectActivity,Wang2022VoidsStability} A mechanism that is valid under one protocol may lose relevance under another. Unless the range of validity is made explicit, mechanistic interpretation cannot reliably support lifetime extrapolation or engineering optimization.\cite{Hering2025AIAccelerated}

\subsection{AI-enabled Approaches for Microscopic Mechanisms of Instability}

AI offers a way to turn these mechanistic challenges into tractable and testable tasks.\cite{Hering2025AIAccelerated,delaAsuncionNadal2025} Its value lies not simply in fitting complex data, but in extracting physically meaningful variables, expanding the accessible scales of simulation, and enabling more systematic links between microscopic events and macroscopic degradation. Chen \textit{et al.} trained a machine-learning interatomic potential for CsPbI$_3$ and used it to map the potential-energy surface associated with perovskite degradation.\cite{Chen2023MLPDegradationCsPbI3} Instead of assuming a single manually prescribed transition path, they resolved several low-barrier pathways from $\gamma$-CsPbI$_3$ toward nonperovskite configurations, showing that instability is controlled by a distribution of accessible pathways rather than by one unique barrier alone.\cite{Chen2023MLPDegradationCsPbI3}

One important role of AI is representation learning. By learning low-dimensional descriptions from atomic structures, local environments, and dynamic trajectories, AI can identify control factors that are difficult to define manually.\cite{Kim2024AllInorganicML,Wang2024CompatibleMolecules,Lee2022ACationScience} This is particularly useful in structurally complex regions such as grain boundaries, surfaces, and interfaces, where multiple variables change simultaneously. When combined with the spatial diagnosis discussed in Section 3, such representations can help identify the shared local motifs of the earliest degraded regions and connect them with measurable spectral or electrical signatures.\cite{Hering2025AIAccelerated} To remain scientifically useful, these learned features should be interpretable and their influence on barriers or rates should be explicitly analyzed.\cite{delaAsuncionNadal2025}

A second major role is played by machine-learning interatomic potentials trained on first-principles data.\cite{Pols2023DefectMigrationMLFF,Liu2023GrainBoundarySlidingMLFF,Samatov2024IonMigrationGBMLFF,Tyagi2025TracingIonMigrationMLFF,Bian2026MLPotentials} These models extend simulation to larger systems and longer times, allowing ion diffusion, defect migration, phase fluctuation, and nucleation behavior to be studied in more realistic microstructures.\cite{Liu2023GrainBoundarySlidingMLFF,Samatov2024IonMigrationGBMLFF} This greatly reduces the gap between atomistic accuracy and mesoscale relevance. For mechanism studies, however, training data must cover the relevant nonequilibrium regions, especially transition states, defective structures, interfaces, and other high-uncertainty configurations. Active learning is therefore particularly important, because it directs data generation toward the parts of configuration space that control the mechanism rather than toward well-behaved equilibrium structures alone.\cite{Tyagi2025TracingIonMigrationMLFF,Hering2025AIAccelerated}

AI is also useful for discovering reaction coordinates and accelerating rare-event sampling.\cite{Chen2022KineticPathwayGammaDelta,Hering2025AIAccelerated} Many mechanistic controversies arise because the actual pathway is not unique. Instead of assuming a single manually chosen path, AI can help identify collective variables and explore distributions of transition pathways under different local environments. This makes mechanistic analysis more realistic and more falsifiable, since it allows one to predict not only a barrier but also the range of possible pathways and intermediate states, together with their expected experimental signatures.\cite{Chen2022KineticPathwayGammaDelta,Tyagi2025TracingIonMigrationMLFF}

To connect microscopic events with larger-scale models, AI can further be used for rapid rate prediction and event-network reduction.\cite{Pols2023DefectMigrationMLFF,Tyagi2025TracingIonMigrationMLFF,Hering2025AIAccelerated} Given a local environment, it can estimate barriers, rate constants, or diffusion coefficients and assign uncertainties. More importantly, it can compress a large event library into a reduced set of dominant pathways and effective parameters that can be passed to kinetic Monte Carlo or continuum-level models. The goal is to identify the pathways that dominate degradation and to quantify the uncertainty associated with their extrapolation.\cite{Tyagi2025TracingIonMigrationMLFF}

Because mechanism-level AI is especially vulnerable to plausible but misleading extrapolations, physical constraints and uncertainty quantification are essential.\cite{Hering2025AIAccelerated,delaAsuncionNadal2025} The models should respect symmetry and conservation requirements, detect out-of-distribution conditions, and remain consistent with multiple forms of evidence rather than a single signal channel. When uncertainty is high, the model should indicate which additional observations or control experiments are most needed, rather than giving overconfident conclusions.\cite{Hering2025AIAccelerated}

\subsection{Validation and Credibility}

The credibility of mechanistic conclusions should be judged by independent validation and by their capacity to generate testable predictions, rather than by fitting quality alone.\cite{Hering2025AIAccelerated,delaAsuncionNadal2025} Key barriers, intermediate states, and reaction motifs predicted by AI-assisted models should be checked against independent first-principles calculations for representative configurations, especially near transition states, interfaces, and defect complexes.\cite{Tyagi2025TracingIonMigrationMLFF,Samatov2024IonMigrationGBMLFF} A convincing mechanism should also explain at least two classes of observation consistently, for example by linking an atomistic migration pathway to correlated electrical drift and spectral evolution.\cite{Wang2024HysteresisDiagnosis,Zhang2022DegradationPathways}

Intervention experiments are equally important. Changes in halide composition, field strength, interface chemistry, or defect passivation should alter the observables in the direction predicted by the proposed mechanism.\cite{Chen2022KineticPathwayGammaDelta,Wang2024CsPbX3Stability} Spatial claims should be checked against the imaging-based statistics discussed in Section 3. If degradation is predicted to begin preferentially at grain boundaries or interfaces, the model should explain why those regions exhibit lower barriers or faster reaction kinetics.\cite{Liu2023GrainBoundarySlidingMLFF,Samatov2024IonMigrationGBMLFF} Finally, the validity range of every model should be stated explicitly, including the protocol window, material morphology, and interface chemistry over which the conclusions are expected to hold. Outside those boundaries, the limits of extrapolation should be made explicit rather than implied.\cite{Hering2025AIAccelerated,delaAsuncionNadal2025}
	
\section{Consequences of Instability}

Sections 3 and 4 established how instability is recognized and why it occurs. The next question is what those instability processes imply for material performance, device lifetime, and reliability.\cite{Li2020OperationalStability,Khenkin2020ConsensusISOS,Tanko2025Reliability,Boyd2019DegradationMechanisms,Wang2022VoidsStability} For CsPbX$_3$, the answer cannot be reduced to a single efficiency-loss curve. Degradation is a stochastic process shaped by coupled mechanisms, stress conditions, and device-to-device variability.\cite{Li2020OperationalStability,Zhang2022DegradationPathways,Tanko2025Reliability} A meaningful reliability description therefore requires not only degradation trajectories, but also explicit failure criteria, statistical dispersion, and uncertainty in extrapolation.\cite{Khenkin2020ConsensusISOS,DunlapShohl2024OperationalLifetimes,Hering2025AIAccelerated} It should also remain connected to the mechanistic parameters discussed in Section 4, so that microscopic information can eventually inform the engineering strategies examined in Section 6.

\subsection{Problem Definition}

The consequences of instability in CsPbX$_3$ appear at several levels.\cite{Li2020OperationalStability,Tanko2025Reliability} At the material level, degradation is reflected in changes in phase composition, defect populations, ion redistribution, interfacial chemistry, and microstructure, including grain evolution, voids, and cracks.\cite{Zhang2022DegradationPathways,Zhou2021DefectActivity,Wang2022VoidsStability} These changes correspond directly to the mechanistic variables discussed in Section 4, but they do not translate automatically into device lifetime. At the device level, instability manifests as the time-dependent degradation of optoelectronic performance, including power conversion efficiency, open-circuit voltage, short-circuit current, fill factor, hysteresis, impedance response, and transient behavior.\cite{Li2020OperationalStability,Chen2016OriginJVHysteresis,Sanchez2014SlowDynamicProcesses} At the system level, the central quantity is the distribution of failure times under specified operating conditions, which defines reliability in the engineering sense.\cite{Khenkin2020ConsensusISOS,Tanko2025Reliability}

This distinction implies that degradation metrics and reliability metrics are not the same.\cite{Khenkin2020ConsensusISOS,Li2020OperationalStability} Degradation metrics describe how a device evolves in time. They include threshold-based quantities such as T80 and T90, fitted parameters such as decay constants or slope changes, and drift rates in electrical or spectral signals.\cite{Khenkin2020ConsensusISOS,DunlapShohl2024OperationalLifetimes} Reliability metrics, by contrast, describe the probability of failure within a given time window. These include lifetime distributions, median or mean lifetime, failure probability, and robustness against fluctuations in operating conditions or manufacturing variation.\cite{Tanko2025Reliability,Hering2025AIAccelerated} A smooth average degradation curve does not necessarily imply high reliability if the device-to-device spread is large. Conversely, a system with only moderate average stability may still be valuable if its behavior is highly reproducible and therefore easier to engineer.\cite{Tanko2025Reliability}

All consequence analysis must remain protocol-aware.\cite{Khenkin2020ConsensusISOS,Zhang2022DegradationPathways,Tanko2025Reliability} In CsPbX$_3$, degradation is often driven by coupled stressors, including temperature, illumination, electrical bias, and environmental atmosphere. Different protocols can change not only the rate of degradation but also the dominant mechanism.\cite{Li2020OperationalStability,Zhang2022DegradationPathways} Some systems show gradual continuous decay, others exhibit long plateaus followed by abrupt failure, and still others show partial reversibility.\cite{Khenkin2020ConsensusISOS,Sanchez2014SlowDynamicProcesses} As a result, degradation curves obtained under different protocols cannot be compared directly unless their mechanistic relationship is made explicit.\cite{Khenkin2020ConsensusISOS,Tanko2025Reliability}

\subsection{Conventional Approaches for Consequences of Instability}

Before the widespread use of AI, the consequences of instability were usually described in a much simpler way.\cite{Li2020OperationalStability,Khenkin2020ConsensusISOS,Boyd2019DegradationMechanisms} Most studies relied on T80 or T90 as the main conclusion, supplemented by a few representative degradation curves.\cite{Khenkin2020ConsensusISOS,DunlapShohl2024OperationalLifetimes} In some cases, empirical fitting functions, such as single or double exponentials and piecewise linear models, were used to compare samples through fitted parameters. When the sample size was small, the best or typical device was often taken as representative of the whole batch, and only a limited number of repeats were reported.\cite{Li2020OperationalStability,Tanko2025Reliability} Zhao \textit{et al.} performed the accelerated-aging study on interface-stabilized all-inorganic CsPbI$_3$ solar cells.\cite{Zhao2022AcceleratedAging} By stressing encapsulated devices under continuous illumination at temperatures up to 110~$^\circ$C, they quantified temperature-dependent degradation trends and showed that once bulk phase stabilization had been improved, interfacial deterioration became a major operational bottleneck.\cite{Zhao2022AcceleratedAging}

More detailed studies sometimes included equivalent-circuit fitting of impedance data or drift--diffusion analysis to interpret hysteresis and bias effects.\cite{Chen2016OriginJVHysteresis,Sanchez2014SlowDynamicProcesses} Even then, however, the link between microscopic mechanism and device lifetime usually remained loose.\cite{Li2020OperationalStability} This traditional approach is simple and intuitive, but it compresses complex degradation into a single number, treats mechanism switching and sudden failure poorly, and provides limited information on statistical variation.\cite{Khenkin2020ConsensusISOS,Tanko2025Reliability} It also offers little support for cross-protocol comparison or reliable extrapolation, since lifetimes measured under different dominant mechanisms are often treated as if they were directly comparable.\cite{Khenkin2020ConsensusISOS,Zhang2022DegradationPathways}

\subsection{Core Challenges in Consequences of Instability}

The main challenge in consequence analysis is not plotting degradation curves, but converting them into actionable reliability knowledge.\cite{Tanko2025Reliability,Hering2025AIAccelerated} One major difficulty is that degradation is often nonlinear and may involve mechanism switching. A device can show rapid early change because of interfacial rearrangement or initial ion migration, followed by an apparently stable plateau, and then abrupt late-stage collapse after phase nucleation or interfacial reaction crosses a critical threshold.\cite{Li2020OperationalStability,Zhang2022DegradationPathways,Wang2022VoidsStability} In such cases, simple time constants or empirical fits have limited physical meaning.\cite{Tanko2025Reliability}

A second challenge is the large and structured dispersion across devices.\cite{DunlapShohl2024OperationalLifetimes,Tanko2025Reliability} Variability may arise from microstructure, defect density, interface quality, impurities, encapsulation, or measurement conditions. Importantly, this spread is often not random noise. It may reflect hidden subpopulations of devices controlled by different degradation pathways. Simple averaging can therefore obscure rather than clarify reliability risk.\cite{Tanko2025Reliability,Hering2025AIAccelerated}

A third challenge concerns extrapolation across protocols.\cite{Khenkin2020ConsensusISOS,Zhang2025OutdoorAgeing,Paraskeva2024OutdoorMonitoring} Accelerated aging tests do not necessarily preserve the same dominant mechanism as real operating conditions. Even modest changes in temperature, illumination, or bias can trigger mechanism switching.\cite{Khenkin2020ConsensusISOS,Li2020OperationalStability} Reliability analysis must therefore identify when extrapolation is justified and when it is fundamentally unsafe.\cite{Tanko2025Reliability,Hering2025AIAccelerated}

A final challenge is the cross-scale connection between mechanism and lifetime. The quantities obtained in Section 4, such as barrier distributions, diffusion coefficients, and reaction rates, describe local events. Translating them into device-level lifetime distributions requires a framework that incorporates field distributions, spatial heterogeneity, and statistical variation.\cite{Hering2025AIAccelerated,delaAsuncionNadal2025} Without such a bridge, mechanism analysis and reliability analysis remain disconnected.

\subsection{AI-enabled Approaches for Consequences of Instability}

AI provides a more flexible framework for consequence analysis by treating degradation as a time-dependent, mechanism-sensitive, and uncertainty-aware process.\cite{DunlapShohl2024OperationalLifetimes,Zhang2024PeLEDLifetime,Hering2025AIAccelerated,delaAsuncionNadal2025} Rather than only fitting a curve after the fact, it can combine early observations, stress conditions, and mechanistic priors to predict later degradation, detect inflection points, and estimate reliability before visible failure occurs.\cite{DunlapShohl2024OperationalLifetimes,Paraskeva2024OutdoorMonitoring} For example, Dunlap-Shohl \textit{et al.} developed physiochemical machine-learning models to predict T$_{80}$ for perovskite solar cells from early electrical behavior and environmental descriptors.\cite{DunlapShohl2024OperationalLifetimes} Their analysis showed that the model could forecast later degradation from partial aging data and also highlighted how temperature, humidity, and photo-oxidative conditions shape operational failure.\cite{DunlapShohl2024OperationalLifetimes} In this sense, AI was used not only as a fitting tool, but as a reliability model that converts short-term observations into lifetime-relevant risk information.

One direct application is degradation trajectory modeling. Once degradation is formulated as a time-series problem, AI can use early-stage spectral, electrical, and performance data to predict short- or medium-term evolution.\cite{DunlapShohl2024OperationalLifetimes,Zhang2024PeLEDLifetime} It can also identify the probability of inflection points, plateau termination, or mechanism switching. This is particularly valuable in engineering practice, because it allows lifetime risk to be estimated from a much shorter observation window and reduces the time needed for screening or process optimization.\cite{DunlapShohl2024OperationalLifetimes,Hering2025AIAccelerated}

A related step is to move from continuous degradation curves to failure-event statistics. Survival-analysis frameworks are especially useful here, because they can incorporate threshold-based failure definitions, handle censored data from unfinished tests, and estimate lifetime distributions together with confidence intervals.\cite{DunlapShohl2024OperationalLifetimes,Hering2025AIAccelerated} More importantly, stress conditions and device descriptors can be introduced as covariates, so that the model does not merely state when failure occurs, but also under which conditions the risk becomes higher.\cite{DunlapShohl2024OperationalLifetimes,Paraskeva2024OutdoorMonitoring} This is particularly relevant for CsPbX$_3$, where abrupt or catastrophic failure may be hidden by average curves but becomes visible in event-based statistics.\cite{Tanko2025Reliability}

When dispersion is large, hierarchical and mixed-effect approaches become necessary.\cite{Hering2025AIAccelerated,delaAsuncionNadal2025} These models can separate within-batch variation from systematic differences between batches, fabrication routes, or device classes. They can also identify latent subgroups of degradation trajectories, such as one subgroup dominated by ion migration and another by interfacial reaction. If the diagnostic outputs of Section 3 are included as covariates or prior labels, the resulting reliability model becomes explicitly mechanism-aware rather than purely statistical.\cite{Hering2025AIAccelerated}

AI is also valuable for mechanism-aware extrapolation. Cross-protocol comparison should not rely on forcing all degradation data into the same empirical formula.\cite{Khenkin2020ConsensusISOS,Hering2025AIAccelerated} Instead, extrapolation should be attempted only when the dominant mechanism remains the same, or when the mapping between mechanisms is understood. In this context, the mechanistic quantities extracted in Section 4, such as barrier distributions, diffusion coefficients, and reaction-rate constants, can serve as physical anchors. They help distinguish genuine acceleration of the same process from a switch to a different degradation pathway. When such a switch is likely, the model should enlarge its uncertainty rather than produce an overconfident lifetime prediction.\cite{DunlapShohl2024OperationalLifetimes,Hering2025AIAccelerated}

Because engineering decisions depend more on risk boundaries than on point estimates, uncertainty quantification is essential.\cite{DunlapShohl2024OperationalLifetimes,Zhang2024PeLEDLifetime,Hering2025AIAccelerated} The model should account for statistical uncertainty arising from noise and device spread, structural uncertainty associated with model form, and extrapolation uncertainty caused by out-of-distribution protocols or fabrication conditions. When uncertainty is high, the model should indicate which additional measurements would reduce it most effectively, thereby connecting directly to the active-learning and closed-loop optimization strategies discussed in Section 6.\cite{Hering2025AIAccelerated,delaAsuncionNadal2025}

\subsection{Validation and Credibility}

Reliability models are especially prone to producing results that fit existing data well but are of limited practical value.\cite{Tanko2025Reliability,Hering2025AIAccelerated} Their credibility should therefore be evaluated using stricter standards. A useful model should be validated not only within the observed time window, but also by using early data to predict later degradation or later failure and then comparing the prediction with subsequent measurements.\cite{DunlapShohl2024OperationalLifetimes,Zhang2024PeLEDLifetime} Validation should also be performed across batches, laboratories, and protocols, since random splits within a single dataset can hide important domain shifts.\cite{Paraskeva2024OutdoorMonitoring,Hering2025AIAccelerated}

Protocol transfer is another critical test.\cite{Khenkin2020ConsensusISOS,Zhang2025OutdoorAgeing,Tanko2025Reliability} The model should be challenged under different temperatures, light intensities, and bias conditions, and any extrapolation boundary should be stated explicitly. When the dominant mechanism changes, the uncertainty should increase accordingly rather than remain artificially small.\cite{DunlapShohl2024OperationalLifetimes,Hering2025AIAccelerated} Reliability outputs such as lifetime distributions, failure probabilities, and confidence intervals must also be calibrated, so that the stated confidence level matches the actual coverage observed in experiment.\cite{Hering2025AIAccelerated,delaAsuncionNadal2025}

Finally, consequence analysis should remain consistent with both diagnosis and mechanism. If a model claims that a particular variable strongly controls lifetime, this conclusion should agree with the mechanistic trends discussed in Section 4 and with the diagnostic evidence of Section 3.\cite{Wang2024HysteresisDiagnosis,DunlapShohl2024OperationalLifetimes} Only under such cross-consistency can reliability modeling serve as a scientifically meaningful bridge between microscopic instability and engineering design.\cite{Hering2025AIAccelerated}
	
\section{Engineering Stability Enhancement}

Sections 3--5 established a unified framework for stability analysis, including transferable diagnosis and early warning, testable microscopic mechanisms, and reliability-oriented consequence modeling.\cite{Hering2025AIAccelerated,delaAsuncionNadal2025} This section turns from analysis to intervention and asks how stability can be improved in real CsPbX$_3$ systems.\cite{Wang2024CsPbX3Stability,Shen2024DefectRegulation,Li2024ModificationStrategies,Xiang2021InorganicStabilityReview,Sun2025OutdoorStableModules,Liang2025BSiteDopingIonMigration} In 3D CsPbX$_3$, stabilization is rarely achieved through a single intervention. It usually requires coordinated control over composition, defects, microstructure, interfaces, processing conditions, and encapsulation, while simultaneously balancing efficiency, lifetime, cost, manufacturability, and reproducibility.\cite{Li2020OperationalStability,Shen2024DefectRegulation,Wang2024CsPbX3Stability} Stability enhancement is therefore best viewed as a constrained multi-objective optimization problem rather than a sequence of isolated improvements.\cite{Hering2025AIAccelerated,Gao2026AutonomousClosedLoop}

\subsection{Problem Definition}

From an engineering perspective, the aim of stabilization is to enhance the ability to reduce failure risk and improve reproducibility across devices and batches.\cite{Khenkin2020ConsensusISOS,Tanko2025Reliability,Hering2025AIAccelerated} A useful strategy should therefore improve median lifetime or performance retention, reduce the probability of premature failure within a given time window, and narrow the spread of degradation behavior. These objectives should be expressed using the reliability metrics introduced in Section 5; otherwise, an apparent lifetime gain may reflect only a few exceptional devices and have limited practical value.\cite{Khenkin2020ConsensusISOS,DunlapShohl2024OperationalLifetimes,Tanko2025Reliability}

The available control variables in 3D CsPbX$_3$ span several levels.\cite{Shen2024DefectRegulation,Li2024ModificationStrategies,Wang2024CsPbX3Stability} At the material level, they include halide composition, mixed-halide ratio, doping, alloying, additive content, and strain regulation.\cite{Swarnkar2016AlphaCsPbI3,Shen2024DefectRegulation,Lee2022ACationScience,Xue2020StrainTransportLayers,Liang2025BSiteDopingIonMigration} At the defect and microstructure level, they include grain size, grain-boundary density, void content, passivation chemistry, and surface modification.\cite{Shen2024DefectRegulation,Li2024ModificationStrategies} At the interface and device level, they include the choice of transport layers, inserted interlayers, contact engineering, ion-blocking structures, and suppression of interfacial reactions.\cite{Zhao2022AcceleratedAging,Shen2024DefectRegulation,Steele2022EmbeddedNetworkCsPbI3,Gong2020BlackPhosphorusQDs} At the processing level, they include solvent systems, crystallization conditions, annealing schedules, post-treatment, encapsulation, and barrier design.\cite{Li2024ModificationStrategies,Laufer2025ProcessMonitoring,Wang2018SolventControlledGrowth} In practice, these variables are strongly coupled, so their effects cannot be assessed reliably through one-factor-at-a-time optimization.\cite{Hering2025AIAccelerated,Gao2026AutonomousClosedLoop}

Any stabilization strategy must also satisfy clear constraints.\cite{Tanko2025Reliability,Hering2025AIAccelerated} It must be manufacturable within a realistic process window, robust against small process fluctuations, and scalable from laboratory-scale devices to larger-area systems.\cite{Khenkin2020ConsensusISOS,Zhao2022AcceleratedAging} Its performance must be evaluated under explicitly defined protocols and statistical standards, and any side effects, such as reduced efficiency or the emergence of new degradation pathways, should be treated as part of the optimization problem rather than ignored.\cite{Khenkin2020ConsensusISOS,Li2020OperationalStability}

\subsection{Conventional Approaches for Engineering Stability Enhancement}

Before the adoption of artificial-intelligence-based methods, stability enhancement in CsPbX$_3$ mainly followed experience-driven and largely modular strategies.\cite{Wang2024CsPbX3Stability,Shen2024DefectRegulation,Li2024ModificationStrategies,Xiang2021InorganicStabilityReview,Sun2025OutdoorStableModules,Liang2025BSiteDopingIonMigration} Composition engineering was used to increase phase stability or raise degradation barriers.\cite{Swarnkar2016AlphaCsPbI3,Wang2024CsPbX3Stability,Wang2018CsPbI3Beyond15} Defect and microstructure control, often achieved through additives, crystallization regulation, and passivation, aimed to suppress nonradiative loss and ion migration.\cite{Shen2024DefectRegulation,Li2024ModificationStrategies} Interface engineering focused on more stable transport layers, interlayers, and contact structures to suppress interfacial reactions and ion penetration.\cite{Zhao2022AcceleratedAging,Shen2024DefectRegulation,Steele2022EmbeddedNetworkCsPbI3,Gong2020BlackPhosphorusQDs} Encapsulation and barrier strategies were then used to reduce the ingress of moisture and oxygen and to improve operational durability.\cite{Khenkin2020ConsensusISOS,Li2020OperationalStability} A representative example is the embedded-interfacial-network strategy of Steele \textit{et al.}, who introduced a PbI$_2$-based interfacial microstructure into otherwise unstable CsPbI$_3$ films and devices.\cite{Steele2022EmbeddedNetworkCsPbI3} They showed that local interfacial design can delay the rapid ambient conversion of black-phase CsPbI$_3$ and thereby improve operational robustness.\cite{Steele2022EmbeddedNetworkCsPbI3}

These approaches have produced many successful examples, but they are often inherently sequential.\cite{Li2024ModificationStrategies,Shen2024DefectRegulation} Improvement at one level frequently shifts the dominant bottleneck to another. For instance, once bulk phase stability is improved, interfacial degradation may become the limiting factor; conversely, passivation may enhance initial efficiency while altering ion migration pathways or long-term reliability.\cite{Zhao2022AcceleratedAging,Wang2024CsPbX3Stability} Traditional optimization therefore tends to operate in a divide-and-conquer mode. It is effective for local improvements, but less effective for systematic search in a high-dimensional space, especially when risk, dispersion, and manufacturability must be considered explicitly.\cite{Hering2025AIAccelerated,Gao2026AutonomousClosedLoop}

\subsection{Core Challenges in Engineering Stability Enhancement}

The practical difficulty of engineering stabilization lies less in generating ideas than in navigating the real design space.\cite{Hering2025AIAccelerated,delaAsuncionNadal2025} The first challenge is the sheer size of the combinatorial space. Composition, additives, process parameters, interface materials, and encapsulation schemes interact nonlinearly, so exhaustive exploration is prohibitively expensive.\cite{Shen2024DefectRegulation,Cakan2025BayesianDurability} The second challenge is the conflict among objectives. Strategies that improve stability may simultaneously compromise charge transport, efficiency, cost, environmental tolerance, or processing simplicity. A solution that looks optimal in one metric may therefore be unusable in practice.\cite{Li2020OperationalStability,Tanko2025Reliability}

A third challenge is that manufacturability defines whether a solution is genuinely useful.\cite{Khenkin2020ConsensusISOS,Gao2026AutonomousClosedLoop} Many strategies work only within very narrow process windows, are sensitive to scale-up, or show large batch-to-batch variation.\cite{Zhao2022AcceleratedAging,Laufer2025ProcessMonitoring} Engineering requires robust solutions rather than extreme cases. A final challenge is the slow and expensive feedback cycle of stability evaluation. Lifetime testing requires time and statistical repetition, and without early proxy indicators or a reliable framework for extrapolation, optimization proceeds very slowly.\cite{DunlapShohl2024OperationalLifetimes,Cakan2025BayesianDurability} This makes it essential to integrate the early-warning outputs of Section 3 and the reliability models of Section 5 into the engineering workflow.\cite{Hering2025AIAccelerated,delaAsuncionNadal2025}

\subsection{AI-enabled Approaches for Engineering Stability Enhancement}

\begin{figure}[tb]
	\begin{center}
		\scalebox{1.5}[1.5]{\includegraphics[width=8cm]{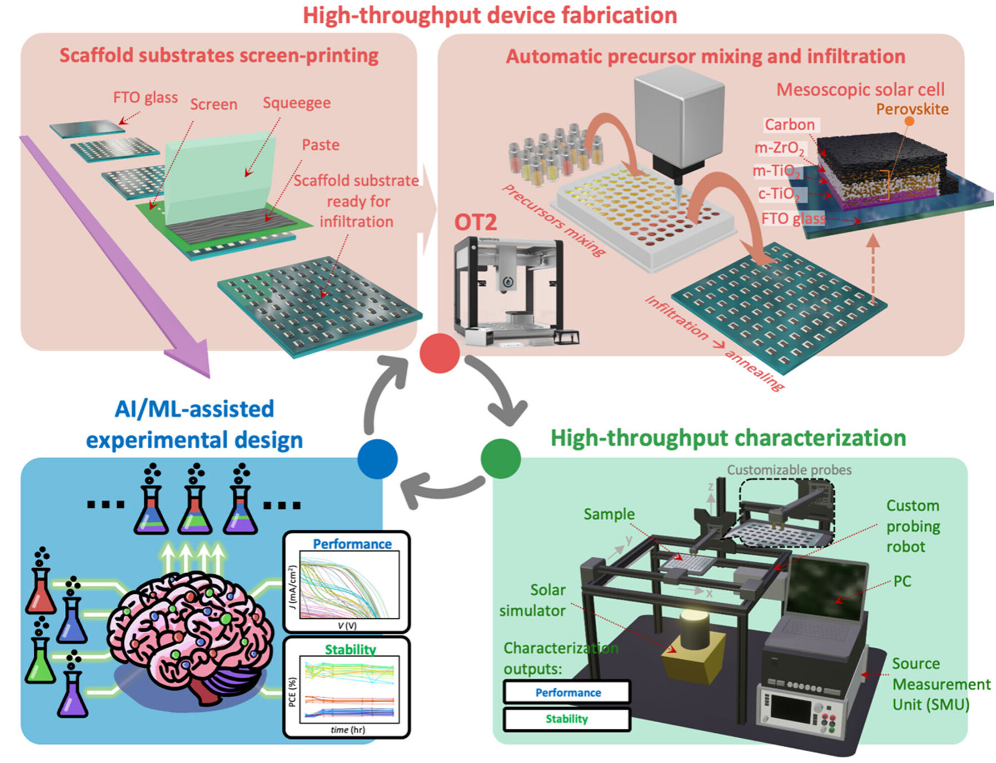}}
	\end{center}
	\caption{(Color online) Schematic of a AI-assisted experimental design for high performance of perovskite solar cell devices.\cite{Miftahullatif2025InDeviceOptimization}}
	\label{fig_ai-optimize}
\end{figure}

AI provides a route to recast stability enhancement from empirical trial-and-error into a closed-loop optimization problem with explicit objectives, constraints, and uncertainty control.\cite{Hering2025AIAccelerated,Cakan2025BayesianDurability,Miftahullatif2025InDeviceOptimization,Gao2026AutonomousClosedLoop,CorreaBaena2018AutomationMLHPC} In this setting, AI is not merely a faster screening tool. Its broader function is to organize prior knowledge, define what should be optimized, predict outcomes from limited data, select the most informative next experiments, and update the design strategy iteratively.\cite{Kim2024AllInorganicML,Wang2024CompatibleMolecules,Gao2026AutonomousClosedLoop} For engineering stabilization, this integrated role is essential because the design space spans composition, additives, processing parameters, interfaces, and encapsulation, while the relevant targets include not only initial efficiency but also lifetime, reproducibility, manufacturability, and robustness. A typical AI-assisted process is shown in Fig.~\ref{fig_ai-optimize} for device designing.\cite{Miftahullatif2025InDeviceOptimization}

A representative AI-guided example is the durability-optimization study of Cakan \textit{et al.}, who used Bayesian optimization to search the composition and additive space of triple-halide perovskite thin films under coupled light and heat stress.\cite{Cakan2025BayesianDurability} They iteratively selected the next most informative experiments and identified films with improved resistance to combined stressors.\cite{Cakan2025BayesianDurability} Miftahullatif \textit{et al.} pushed this logic closer to device manufacturing by coupling machine learning with an all-printed fabrication workflow, showing that only a small fraction of the accessible process space needed to be sampled to improve device metrics.\cite{Miftahullatif2025InDeviceOptimization} These studies make the role of AI concrete, i.e., stability enhancement is reformulated as a sequential decision problem in which each experiment updates the next one.

Natural-language-based large language models (LLMs) are particularly useful at the knowledge-organization stage, because much of the experience relevant to stabilization is stored not in standardized numerical tables, but in papers, patents, processing notes, and experimental recipes. In perovskite research, these models can serve as literature-to-structure interfaces. They extract additives, processing parameters, interface materials, degradation stressors, and performance outcomes from text, and then convert them into structured datasets or knowledge graphs that can be queried by downstream optimization models.\cite{Xie2024StructuredDatasetLLM,Sipila2025QAModels,Liu2025PerovskiteLLM,Li2026KnowledgeReconstructionLLM} This is especially relevant for stability engineering, where valuable prior knowledge is often qualitative and highly dispersed across the literature, for example, which additives suppress ion migration, which Lewis bases regulate crystallization, or which interlayers mitigate interfacial reaction. Instead of leaving such information as unstructured background knowledge, LLM pipelines can transform it into machine-readable constraints, ranked candidate lists, and mechanistic hypotheses that define a more realistic search space for subsequent optimization.

Domain-specialized perovskite LLMs further extend this role from passive information retrieval to active design support. Perovskite-LLM combines a perovskite knowledge graph, instruction-tuning datasets, and specialized reasoning models to support domain-specific question answering and scientific reasoning, including literature review and experimental design.\cite{Liu2025PerovskiteLLM} Perovskite-R1 moves one step further toward actionable discovery by targeting precursor-additive design. After mining high-quality publications and integrating a large candidate-material library, it proposes additive strategies and experimental plans that were then tested experimentally.\cite{Wang2026PerovskiteR1} Although these demonstrations are centered mainly on perovskite solar cells rather than 3D all-inorganic CsPbX$_3$ specifically, the same logic is directly transferable to instability engineering in inorganic perovskites. In practice, an LLM could be used to prioritize inorganic-compatible additives, surface ligands, interfacial modifiers, dopants, or encapsulation chemistries under explicit constraints of phase stability, ion-migration suppression, processing temperature, and compatibility with CsPbI$_3$/CsPbBr$_3$ fabrication windows. In this role, the LLM should not replace the surrogate model or Bayesian optimizer; it is better understood as a knowledge engine that narrows the search space before expensive lifetime tests are launched.

A further opportunity is to couple LLMs with closed-loop optimization and autonomous experimentation.\cite{Chaudhari2026MatSciAgent,Sun2026PVKLLM,Mandal2025AILAAFM} Recent work in perovskite and broader materials science shows that language models can bridge natural-language goals, structured recipes, simulation tools, and laboratory actions, thereby helping initialize Bayesian optimization, translate expert heuristics into machine-executable protocols, and update the next experiment using both literature priors and new feedback.\cite{Sun2026PVKLLM,Chaudhari2026MatSciAgent} For stability enhancement, this is valuable when the objective is not simply to maximize initial efficiency, but to optimize multiple targets simultaneously, such as lifetime, retention under thermal/light/bias stress, batch reproducibility, and manufacturability. An LLM-guided workflow could, for instance, propose a small batch of chemically plausible additive or interface candidates, explain which degradation channel each candidate is expected to suppress, and then pass these candidates to active-learning or Bayesian-optimization loops for experimental verification.

The limitations of LLM-based workflows, however, must be stated clearly. Their outputs are strongly shaped by literature bias, reporting inconsistency, and hallucination risk, and good performance on question-answer benchmarks does not automatically translate into reliable laboratory execution.\cite{Mandal2025AILAAFM,Li2026KnowledgeReconstructionLLM} For instability research, where conclusions are highly protocol-dependent, LLM recommendations should therefore be retrieval-grounded, linked to structured metadata, and validated by physics-based modeling together with targeted experiments, rather than treated as autonomous scientific conclusions.\cite{Wang2026PerovskiteR1,Liu2025PerovskiteLLM,Sun2026PVKLLM} Under these conditions, natural-language models can serve as a practical front end for stability engineering by summarizing dispersed knowledge and generating constrained hypotheses that improve the efficiency of downstream optimization.

Once prior knowledge has been organized, a key next step is to formalize the engineering objective. Stability improvement should be written as a computable target, such as median lifetime, failure probability within a target window, degradation inflection time, or risk-adjusted performance retention.\cite{DunlapShohl2024OperationalLifetimes,Hering2025AIAccelerated} These objectives can be combined with efficiency and other functional metrics, while manufacturability, cost, dispersion, and robustness are imposed as constraints. This prevents the search from converging toward isolated high-performing cases that cannot be reproduced or scaled.\cite{Tanko2025Reliability,Hering2025AIAccelerated}

Because the design space is large and each evaluation can be expensive, surrogate models are especially useful at the next stage.\cite{Cakan2025BayesianDurability,Miftahullatif2025InDeviceOptimization,Gao2026AutonomousClosedLoop} These models approximate the mapping from design variables to reliability-oriented outputs while providing uncertainty estimates. Their inputs may include composition, additive chemistry, processing parameters, interface descriptors, and early diagnostic signals from Section 3 as fast proxy indicators. Their outputs should not be limited to a single lifetime value, but should include the reliability metrics and confidence bounds discussed in Section 5.\cite{DunlapShohl2024OperationalLifetimes,Cakan2025BayesianDurability} To remain useful, such models must also be protocol-aware and calibrated against domain shifts.\cite{Hering2025AIAccelerated,delaAsuncionNadal2025}

With a surrogate model in place, active learning and Bayesian optimization can be used to allocate experimental resources more efficiently.\cite{Cakan2025BayesianDurability,Wang2024CompatibleMolecules,Gao2026AutonomousClosedLoop,CorreaBaena2018AutomationMLHPC} Instead of sweeping parameters uniformly, the next experiment is selected to maximize expected improvement or information gain. This enables the search to move strategically between exploration of uncertain regions and exploitation of promising regions, while avoiding unnecessary testing of clearly poor candidates.\cite{Cakan2025BayesianDurability} The essential point is that each round of experiments should either improve the target objective or reduce the uncertainty that currently limits decision-making.\cite{Hering2025AIAccelerated,Gao2026AutonomousClosedLoop}

Robustness improves further when the optimization is made mechanism-aware and supported by early proxy indicators.\cite{Hering2025AIAccelerated,delaAsuncionNadal2025,DunlapShohl2024OperationalLifetimes,Cakan2025BayesianDurability} Black-box optimization based only on recipes and process conditions often suffers from limited transferability. A more reliable strategy is to introduce mechanistic quantities from Section 4, such as migration-barrier distributions, interfacial reactivity indicators, or defect-related descriptors, as intermediate targets or auxiliary constraints.\cite{Kim2024AllInorganicML,Wang2024CompatibleMolecules,Lee2022ACationScience} At the same time, early-warning signals identified in Section 3 can be used to screen out obviously high-risk candidates before expensive long-duration tests are performed. With a smaller set of calibrated lifetime experiments, these early indicators can then be linked quantitatively to lifetime distributions. In this way, the optimization cycle shifts from one based entirely on slow endpoint testing to one based on rapid screening followed by targeted validation.\cite{Cakan2025BayesianDurability,Hering2025AIAccelerated}

When these elements are connected to automated synthesis and testing, AI naturally leads to closed-loop experimental platforms.\cite{Laufer2025ProcessMonitoring,Gao2026AutonomousClosedLoop,CorreaBaena2018AutomationMLHPC} Even partial automation can improve metadata collection, protocol consistency, anomaly screening, and model updating, thereby reducing human variability and increasing reproducibility. Full automation is not required for this framework to be useful; even semi-automated pipelines can substantially strengthen the rigor and efficiency of stability optimization.\cite{Laufer2025ProcessMonitoring,Hering2025AIAccelerated} In this overall framework, LLMs provide literature-grounded priors and constrained hypotheses, surrogate models and Bayesian optimization decide where to search next, and experiments provide the feedback needed to recalibrate the model. Stability enhancement then becomes a continuously updated decision process rather than a sequence of isolated empirical fixes.

\subsection{Validation and Credibility}

A stabilization strategy should be considered effective only when it satisfies reliability-oriented standards rather than isolated device-level improvements.\cite{Khenkin2020ConsensusISOS,Tanko2025Reliability} Statistical validation is essential. Lifetime distributions and failure probabilities should be reported with sufficient sample size, and batch effects should be made explicit.\cite{DunlapShohl2024OperationalLifetimes,Tanko2025Reliability} Robustness must also be tested under perturbations of the target protocol, including moderate changes in temperature, illumination, bias, or humidity, to determine whether the risk remains controlled.\cite{Khenkin2020ConsensusISOS,Zhao2022AcceleratedAging}

Cross-scale validation is equally important.\cite{Tanko2025Reliability,Hering2025AIAccelerated} A strategy that works on a small-area device must be tested for consistency when the area, substrate, or device architecture changes. In addition, the dominant failure mode should be tracked. A strategy that merely delays one degradation pathway but shifts the system to a different failure mechanism cannot be regarded as fully validated until the new mode is also characterized within the reliability framework.\cite{Li2020OperationalStability,Zhao2022AcceleratedAging} Finally, every strategy should include a clear statement of its applicability range, especially when it depends on specific interfaces, process conditions, or encapsulation schemes.\cite{Khenkin2020ConsensusISOS,Tanko2025Reliability}
	
\section{Boundaries for AI Methods}

The previous sections showed how AI can assist diagnosis, mechanistic analysis, reliability modeling, and engineering optimization in CsPbX$_3$ instability research.\cite{Hering2025AIAccelerated,delaAsuncionNadal2025,Tao2021MLPerovskiteDesign,CorreaBaena2018AutomationMLHPC} None of these applications, however, is credible without reliable data, meaningful benchmark tasks, and reproducible workflows.\cite{Jacobsson2022FAIRDatabase,Unger2022PerovskiteDatabase,Hansen2022FAIRness} This section therefore turns from capability to boundary conditions. The focus is on the data foundations of AI-assisted stability research, with particular attention to heterogeneity, protocol dependence, reproducibility, and the practical limits of current methods.\cite{Maqsood2025InteroperablePerovskite,Hering2025AIAccelerated,delaAsuncionNadal2025}

\subsection{Data Curation}

Stability research data are inherently heterogeneous.\cite{Chakraborty2023CuratedData,Jacobsson2022FAIRDatabase,Maqsood2025InteroperablePerovskite,Tao2021MLPerovskiteDesign} They may include microscopy images, spectral measurements, time-series electrical signals, tabulated material properties, structural models, first-principles calculations, molecular-dynamics trajectories, and device-level lifetime tests. These data are generated under widely varying experimental conditions, including different temperatures, humidities, light intensities, electrical biases, materials, fabrication processes, and device architectures.\cite{Jacobsson2022FAIRDatabase,Hering2025AIAccelerated} Computational data add another layer of variability because they depend on model choice, boundary conditions, and potential accuracy. This heterogeneity makes data integration difficult and can easily produce misleading comparisons if protocol differences are ignored.\cite{Chakraborty2023CuratedData,Maqsood2025InteroperablePerovskite}

For this reason, data quality control is a prerequisite rather than a downstream refinement.\cite{Hansen2022FAIRness,Chakraborty2023CuratedData} Experimental noise, measurement drift, missing entries, inconsistent units, and incomplete reporting all affect the reliability of subsequent models. Cleaning, normalization, anomaly screening, and format unification are therefore central scientific requirements rather than technical housekeeping.\cite{Jacobsson2022FAIRDatabase,Unger2022PerovskiteDatabase} This is especially important when combining datasets from different laboratories, because missing metadata or inconsistent reporting can make apparently similar data fundamentally incomparable.\cite{Hansen2022FAIRness,Maqsood2025InteroperablePerovskite}

Standardization is the durable solution.\cite{Jacobsson2022FAIRDatabase,Maqsood2025InteroperablePerovskite,Hering2025AIAccelerated} Stability datasets should include clearly defined metadata describing stress protocols, sample and device structures, measurement methods, instrumentation, and data-processing procedures.\cite{Jacobsson2022FAIRDatabase,Unger2022PerovskiteDatabase} Protocol definitions for temperature, humidity, illumination, bias, and sampling intervals should be explicit, as should the conventions used for quantities such as T80 and T90.\cite{Khenkin2020ConsensusISOS,Jacobsson2022FAIRDatabase} Without this degree of standardization, even large datasets may fail to support meaningful cross-study comparison or transferable models.\cite{Hansen2022FAIRness,Maqsood2025InteroperablePerovskite,delaAsuncionNadal2025}

\subsection{Benchmarks and Task Definitions}

Once the data foundation is in place, benchmark tasks and evaluation standards must be defined clearly.\cite{Hering2025AIAccelerated,delaAsuncionNadal2025} In stability research, benchmark tasks may include failure-mode classification, degradation-trajectory prediction, lifetime-distribution modeling, or protocol transfer. Their purpose is not merely to compare algorithms, but to test whether models solve scientifically meaningful problems under realistic conditions.\cite{delaAsuncionNadal2025}

Evaluation should therefore go beyond standard metrics such as accuracy, F1 score, root mean square error, or area under the receiver operating characteristic curve.\cite{delaAsuncionNadal2025,Sipila2025QAModels} These measures remain useful, but they should be supplemented by criteria that matter directly for stability research, such as the accuracy of degradation forecasting, the calibration of lifetime distributions, the robustness of predictions across protocols, and the transferability of the model across laboratories.\cite{Hering2025AIAccelerated,delaAsuncionNadal2025} A benchmark that performs well only under random train--test splits within a single dataset has limited scientific value if it fails under the domain shifts encountered in real applications.\cite{Hering2025AIAccelerated,Maqsood2025InteroperablePerovskite}

Benchmark construction also requires carefully curated datasets.\cite{Chakraborty2023CuratedData,Jacobsson2022FAIRDatabase,Sipila2025PV600Dataset} Such datasets should cover representative materials, stress conditions, fabrication routes, and measurement types, and each record should contain standardized labels for composition, protocol, and outcome.\cite{Jacobsson2022FAIRDatabase,Xie2024StructuredDatasetLLM} Public accessibility is particularly important, because benchmark value increases only when different groups can test models under the same conditions and compare results transparently.\cite{Unger2022PerovskiteDatabase,Hansen2022FAIRness} Open benchmark datasets therefore serve not only as evaluation tools, but also as a mechanism for community-wide convergence on best practice.\cite{Jacobsson2022FAIRDatabase,Sipila2025PV600Dataset}

\subsection{Reproducibility and Transparency}

Reproducibility is a central requirement in AI-assisted stability research because complex models can easily produce results that look impressive yet remain difficult to verify independently.\cite{Hering2025AIAccelerated,delaAsuncionNadal2025} Reproducibility depends not only on access to code, but also on transparency in data sources, preprocessing steps, model settings, and training procedures. Without such transparency, even nominally open studies may remain effectively irreproducible.\cite{Hansen2022FAIRness,Unger2022PerovskiteDatabase}

Data transparency is especially important because stability datasets often involve subjective choices in data selection, labeling, and filtering.\cite{Chakraborty2023CuratedData,Sipila2025PV600Dataset} Experimental design, data-processing steps, and metadata definitions should therefore be made explicit.\cite{Jacobsson2022FAIRDatabase,Maqsood2025InteroperablePerovskite} Wherever possible, both data and code should be released in a form that allows independent groups to rerun the analysis under the same conditions.\cite{Jacobsson2022FAIRDatabase,Unger2022PerovskiteDatabase,Sipila2025PV600Dataset} Reproducibility is strengthened further when multiple groups validate the same model using shared benchmark datasets and consistent reporting conventions.\cite{Hansen2022FAIRness,Hering2025AIAccelerated}

Uncertainty management is another essential component of reproducibility.\cite{Hering2025AIAccelerated,delaAsuncionNadal2025} Models should provide not only point predictions but also confidence estimates, particularly when they are used for extrapolation or decision-making. This is crucial in stability research, where noise, protocol shifts, and limited sample size are common. A model that reports uncertainty honestly is often more scientifically useful than one that reports only a seemingly precise but weakly justified prediction.\cite{Hering2025AIAccelerated,delaAsuncionNadal2025}

\subsection{Boundaries of AI Methods}

Although AI offers powerful tools for stability research, its limitations should be stated plainly.\cite{delaAsuncionNadal2025,Hering2025AIAccelerated} Model performance depends strongly on the quantity, quality, and representativeness of the training data. When data are sparse, biased, or poorly standardized, models may fail to learn meaningful patterns.\cite{Chakraborty2023CuratedData,Maqsood2025InteroperablePerovskite} Overfitting remains a persistent risk, especially when models are evaluated only within a narrow dataset.\cite{delaAsuncionNadal2025} In addition, high predictive accuracy does not automatically imply physical understanding. Many models remain weak in interpretability, which limits their value for mechanism inference and scientific extrapolation.\cite{Hering2025AIAccelerated,delaAsuncionNadal2025}

For these reasons, AI should be treated as a complement to physical modeling and experiment rather than as a replacement for them.\cite{Hering2025AIAccelerated,delaAsuncionNadal2025} Its most useful role is to help researchers organize complex data, identify patterns, prioritize hypotheses, and guide experimental design.\cite{Sipila2025QAModels,Xie2024StructuredDatasetLLM,Laufer2025ProcessMonitoring} The strongest results usually come from combining AI with mechanistic insight, well-defined protocols, and careful experimental validation.\cite{Hering2025AIAccelerated,Hansen2022FAIRness,CorreaBaena2018AutomationMLHPC,Bian2026MLPotentials}

\section{Conclusion and Outlook}
	
This review has examined the instability of 3D CsPbX$_3$ from the linked perspectives of diagnosis, microscopic mechanism, reliability, and engineering optimization, with AI embedded throughout the workflow as summarized in Table~\ref{tab_summary}. The main conclusion is that AI is most valuable as a bridge connecting multimodal diagnosis, mechanism discovery, reliability analysis, and stability-oriented engineering design.
	
A central message of the review is that instability research should be treated as a connected problem rather than as a set of disconnected tasks. Diagnosis must identify degradation modes and early signals. Mechanistic analysis must explain why these processes occur and under what conditions they dominate. Reliability modeling must translate degradation into lifetime distributions, failure probabilities, and uncertainty-aware risk metrics. Engineering optimization must then use these outputs to search for stabilization strategies that are not only effective, but also robust and manufacturable. AI can strengthen each of these steps and, more importantly, can tighten the links among them.
	
The long-term value of AI in this field will depend on foundations that are often less visible but ultimately more decisive. These include protocol-aware datasets, transparent metadata standards, meaningful benchmark tasks, reproducible workflows, calibrated uncertainty estimates, and interpretable cross-scale models. Without these foundations, even highly accurate models may remain difficult to trust, compare, or reuse.
	
Looking forward, several directions appear especially important. The first is the construction of standardized data infrastructures that support comparison across laboratories and protocols. The second is the development of interpretable models that connect atomistic mechanisms with device-level reliability. The third is the wider adoption of closed-loop workflows that integrate AI with automated or semi-automated experiments. The fourth is the establishment of engineering-oriented standards that treat stability together with efficiency, manufacturability, reproducibility, and cost.
	
In short, AI has already become a useful tool for studying inorganic-perovskite instability. Its broader impact, however, will depend on whether it can be embedded in a framework that is physically grounded, experimentally verifiable, statistically reliable, and practically useful. If these conditions are met, AI will play an increasingly important role in the rational design of more stable inorganic perovskites for future optoelectronic and energy applications.
	
\textbf{Acknowledgments} This work was supported by the National Natural Science Foundation of China (Grant Nos. 12072182 and 12421002).
	
	
\end{document}